\documentclass[fleqn,12pt]{article}
\usepackage{cite}
\usepackage{amsmath}
\usepackage{amssymb}
\usepackage{amsfonts}
\usepackage{bbm} 
\usepackage{graphicx} 
\usepackage{subfigure}
\usepackage[small]{caption2}
\setcaptionmargin{1cm}

\DeclareMathOperator{\tr}{tr}
\DeclareMathOperator{\diag}{diag}
\DeclareMathOperator{\ad}{ad}

\def\I{\mathrm{i}}
\newcommand{\SimpleRoot}[1]{\alpha_{(#1)}{}}
\newcommand{\FundamentalWeight}[1]{\mu^{(#1)}}
\newcommand{\CenterObject}[1]{\ensuremath{\vcenter{\hbox{#1}}}}
\renewcommand{\thesection}{\arabic{section}}
\renewcommand{\thesubsection}{\arabic{section}.\arabic{subsection}}
\renewcommand{\thesubsubsection}{(\roman{subsubsection})}
\numberwithin{equation}{section}
\numberwithin{table}{section}
\renewcommand{\thetable}{\arabic{section}.\arabic{table}}
\renewcommand{\theequation}{\arabic{section}.\arabic{equation}}

\allowdisplaybreaks[1]
\begin{document}
\thispagestyle{empty}
\noindent
DESY 03-069 \hspace*{\fill} June 05, 2003\\
\vspace*{1.6cm}

\begin{center}
{\Large\bf Group-Theoretical Aspects of\\[.3cm] Orbifold and Conifold GUTs}
\\[2.0cm]
{\large Arthur Hebecker and Michael Ratz}\\[.5cm]
{\it Deutsches Elektronen-Synchrotron, Notkestrasse 85, D-22603 Hamburg,
Germany}
\\[1cm]

{\bf Abstract}\end{center}
\noindent
Motivated by the simplicity and direct phenomenological applicability 
of field-theoretic orbifold constructions in the context of 
grand unification, we set out to survey the immensely rich 
group-theoretical possibilities open to this type of model 
building. In particular, we show how every maximal-rank, regular subgroup 
of a simple Lie group can be obtained by orbifolding and determine under 
which conditions rank reduction is possible. We investigate how standard 
model matter can arise from the higher-dimensional SUSY gauge multiplet. 
New model building options arise if, giving up the global orbifold 
construction, generic conical singularities and generic gauge twists 
associated with these singularities are considered. Viewed from 
the purely field-theoretic perspective, such models, which one 
might call conifold GUTs, require only a very mild relaxation of the 
constraints of orbifold model building. Our most interesting concrete 
examples include the breaking of E$_7$ to SU(5) and of E$_8$ to 
SU(4)$\times$SU(2)$\times$SU(2) (with extra factor groups), where three 
generations of standard model matter come from the gauge sector and the 
families are interrelated either by SU(3) R-symmetry or by an SU(3) 
flavour subgroup of the original gauge group.

\newpage

\section{Introduction}
Arguably, the way in which fermion quantum numbers are explained by 
\(\mathrm{SU}(5)\)-related grand unified theories (GUTs) represents one of 
the most profound hints at fundamental physics beyond the standard model 
(SM)~\cite{gg,gfm} (also~\cite{ps}). In this context supersymmetry (SUSY), 
usually invoked to solve the hierarchy problem and to achieve gauge coupling 
unification, receives a further and maybe even more fundamental motivation: 
If the underlying gauge group contains gauge bosons with the quantum numbers 
of SM matter, SUSY enforces the existence of the corresponding 
fermions. This is very naturally realized in a higher-dimensional setting, 
where the extra-dimensional gauge field components and their fermionic 
partners can be light even though the gauge group is broken at a high scale
(see, e.g.,~\cite{Babu:2002ti,wy,gmn,nb}). 

Thus, we adopt the point of view that, at very high energies, we are 
faced with a super Yang-Mills (SYM) theory in $d>4$ dimensions which is 
compactified in such a way that the resulting 4d effective theory has 
smaller gauge symmetry (ideally that of the SM) and contains the light 
SM matter and Higgs fields. In the simplest models the compactification
space is flat except for a finite number of singularities. Although this 
situation arises naturally in heterotic string theory~\cite{wit,dhvw} and 
has thus been extensively studied in string model building, it has only 
recently been widely recognized that many interesting phenomenological 
implications do not depend on the underlying quantum gravity model and can 
be studied directly in higher-dimensional field theory~\cite{kaw} 
(also~\cite{af,hn,hm,abc,hnos}). 

In the purely field-theoretic context, one has an enormous freedom in
choosing the underlying gauge group, the number of extra dimensions and 
their geometry, the way in which the compactification reduces the gauge 
symmetry (e.g., the type of orbifold breaking), the possible extra field 
content and couplings in the bulk and at the singularities. Although, 
using all this freedom, realistic models can easily be constructed, there 
is so far no model which, by its simplicity and direct relation to the 
observed field content and couplings, appears to be as convincing as, say, 
the generic SU(5) unification idea. However, we feel that the search for
such a model in the framework of higher-dimensional SYM theory is 
promising and that a thorough understanding of the group-theoretical 
possibilities of orbifold-breaking (without the restrictions of 
string theory) will be valuable in this context. The present paper is 
aimed at the exploration of these possibilities and their application 
to orbifold GUT model building. In particular, we are interested in methods 
for breaking larger gauge groups to the SM, in possibilities 
for rank reduction, and in the derivation of matter fields from the adjoint 
representation. 

In Sec.~\ref{gt}, we collect some of the most relevant facts 
and methods of group theory, which serves in particular to fix our notation
and conventions for the rest of the paper. 

In Sec.~\ref{sg}, we begin by recalling the generic features of field 
theoretic orbifold models. It is then shown how orbifolding can break a 
simple Lie group to any of its maximal regular subgroups. This implies, in 
particular, that any regular subgroup (possibly times extra 
simple groups and U(1) factors) can be obtained by orbifold-breaking and 
opens up an enormous variety of model building possibilities. 
 
We continue in Sec.~\ref{rr} by exploring rank reduction by non-Abelian 
orbifolding. We show that simple group factors can always be broken 
completely. In cases where a maximal subgroup contains an extra U(1) factor, 
this factor can only be broken under certain conditions. We give a 
criterion specifying when the extra U(1) cannot be removed. As an interesting 
observation, we note that under special circumstances rank reduction based
on inner automorphisms is also possible on Abelian orbifolds.

In Sec.~\ref{coni}, we discuss manifolds with conical singularities which 
can not be obtained by orbifolding. In particular, such `conifolds' can 
have conical singularities with arbitrary deficit angle. In addition, we
consider the possibility of having Wilson lines with unrestricted values 
wrapped around the singularities of orbifolds or conifolds. All this 
gives rise to many new possibilities for gauge symmetry breaking and for the 
generation of three families of chiral matter from the field content of the 
SYM theory. 

Finally, Sec.~\ref{mo} discusses three specific models, one with E$_7$ 
broken to SU(5) and two with E$_8$ broken to SU(4)$\times$SU(2)$\times$SU(2) 
(with extra factor groups). In all cases, three generations of SM matter 
come from the gauge sector. In one of the E$_8$ models, the 
families are interrelated by an SU(3) R-symmetry, while in the two other 
models an SU(3) flavour subgroup of the original gauge group appears. 

Sec.~\ref{co} contains our conclusions and outlines future perspectives 
and open questions. 

\section{Basics of group theory}\label{gt}

This section is not meant as an introduction to group theory, but merely 
serves to remind the reader of some crucial facts and to fix our notation. 
Relevant references include the classic papers of Dynkin~\cite{dyn1,dyn2,
dyn3} (partially collected in~\cite{dynb}), various textbooks 
(e.g.,~\cite{gil,Georgi:1999jb,cahn}), and the review article~\cite{sla}. 

For each finite-dimensional, complex Lie algebra \(\mathfrak{g}\), the 
maximal Abelian subalgebra \(\mathfrak{h}\), which is unique up to 
automorphisms, is called Cartan subalgebra. 
Its dimension defines the rank $r$ of the Lie algebra and 
its generators will be denoted by \(\{\boldsymbol{H}_i\}_{i=1}^r\). 
They are orthonormal with respect to the Killing metric, i.e.,
they fulfill the relation
\begin{equation}
 \tr (\boldsymbol{H}_i\,\boldsymbol{H}_j)
 \,=\,\lambda\,\delta_{ij}\;,\label{norm}
\end{equation}
where the trace is taken in the adjoint representation and
\(\lambda\) is some constant.

The remaining generators can be chosen such that
\begin{equation}
 [\boldsymbol{H}_i,\boldsymbol{E}_{\alpha}]
 \,=\,
 \alpha_i\,\boldsymbol{E}_{\alpha}\;,\label{hec}
\end{equation}
and are called roots. They are normalized as in Eq.~(\ref{norm}). Each 
root \(\boldsymbol{E}_{\alpha}\) is determined uniquely by the root 
vector \(\alpha\), which is an element of an $r$-dimensional Euclidean 
space, called the root space. The set of all roots will be denoted 
by \(\Sigma\). The \(\boldsymbol{E}_\alpha\) obey the commutation relations
\begin{equation}
 \left[\boldsymbol{E}_\alpha,\boldsymbol{E}_\beta\right]
 \,=\,
 N_{\alpha,\beta}\,\boldsymbol{E}_{\alpha+\beta}\;,
 \label{eq:DefOfNalphabeta}
\end{equation}
where the \(N_{\alpha,\beta}\) are normalization constants,
and \(N_{\alpha,\beta}=0\) means that \(\alpha+\beta\not\in\Sigma\).

We introduce an order in the root space by
\begin{equation}
 \alpha-\beta >0
 \quad:\Leftrightarrow\quad
 \text{first non-vanishing component of}\:\alpha-\beta>0
 \;.
\end{equation}
Correspondingly, we will call a root `positive' if the first non-vanishing
component in the root basis is positive.
The smallest \(r\) positive roots are called simple and will be denoted
by \(\{\SimpleRoot{i}\}_{i=1}^r\).
They are linearly independent, and any root can be expressed by a linear
combination
\begin{equation}
 \alpha = \sum\limits_{i=1}^r k^i\,\SimpleRoot{i}
\end{equation}
with integer coefficients \(k^i\). Motivated by this, a basis
\begin{equation}
 e_{(i)}
 \,=\,
 \frac{2}{|\alpha_{(i)}|^2}\SimpleRoot{i}\;,\label{eb}
\end{equation}
is introduced. The normalization factor will be justified later.

In this basis, the Euclidean metric of the root space is characterized by 
\(g_{ij}=e_{(i)}\cdot e_{(j)}\). It is useful to consider also the vector 
space dual to the root space which, given the existence of a metric in the 
root space, can be identified with the root space by the canonical
isomorphism. It is spanned by the so-called fundamental weights
\(\FundamentalWeight{i}\) (\(1\le i\le r\)) which are defined by
\begin{equation}
 \mu^{(i)}\cdot e_{(j)} \,=\,\delta^i_j\;.
\end{equation}
The components with respect to the $\mu^{(i)}$ basis are called Dynkin 
labels. Correspondingly, the $\mu^{(i)}$ are frequently referred to as the 
Dynkin basis, in which case the $e_{(i)}$ are called the dual basis. 
The constant \(\lambda\) in Eq.~\eqref{norm} 
is chosen such that \(|\SimpleRoot{i}|^2=2\) for the longest of the simple
roots. Then the normalization factor in Eq.~(\ref{eb}) ensures that the 
Dynkin label of any weight (weights being the analogues of the vectors 
$\alpha$ in an arbitrary representation) is integer valued.

The Dynkin labels of each simple root are given by the corresponding row
of the Cartan-matrix
\begin{equation}
 A_{ij}
 \,=\,
 2\,\frac{\SimpleRoot{i}\cdot\SimpleRoot{j}}{|\SimpleRoot{j}|^2}\,=\,
 g_{ij}\,\frac{|\SimpleRoot{i}|^2}{2}\;,
\end{equation}
which encodes the metric of the root space. 

It is well-known that there exist four infinite series of simple
groups \(\mathrm{A}_r\), \(\mathrm{B}_r\), \(\mathrm{C}_r\) 
and \(\mathrm{D}_r\), corresponding to the classical groups,
and the exceptional groups \(\mathrm{G}_2\), \(\mathrm{F}_4\), 
\(\mathrm{E}_6\), \(\mathrm{E}_7\) and \(\mathrm{E}_8\).
The scalar products of the simple roots determine the Dynkin diagrams
(cf.\ the captions of Tab.~\ref{tab:DynkinDiagramsOfClassicalGroups}
and Tab.~\ref{tab:DynkinDiagramsOfExceptionalGroups}).

For later convenience, we introduce the most negative root \(\theta\), 
which leads us to the extended Dynkin diagrams as listed in 
Tab.~\ref{tab:DynkinDiagramsOfClassicalGroups} and in 
Tab.~\ref{tab:DynkinDiagramsOfExceptionalGroups}.

\begin{table}[ht]
 \begin{center}
  \begin{tabular}{|c|l|l|}
   \hline
   \multicolumn{1}{|c}{Name} & 
   \multicolumn{1}{|c}{Real algebra} & 
   \multicolumn{1}{|c|}{Extended Dynkin diagram} \\
   \hline 
   & &  \\[-3mm]
   \(\mathrm{A}_n \) & \( \mathfrak{su}(n+1)\) 
   & \(\CenterObject{\vphantom{\includegraphics[scale=1.1]
   {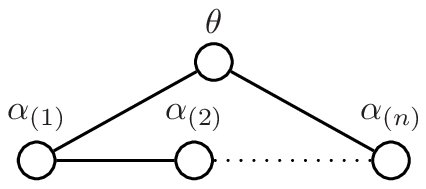}}\includegraphics{DynkinAnExtendedLabeled.eps}
   }\) \\
   & &  \\
   \(\mathrm{B}_n \) & \(  \mathfrak{so}(2n+1)\) 
   & \(\CenterObject{\vphantom{\includegraphics[scale=1.1]
   {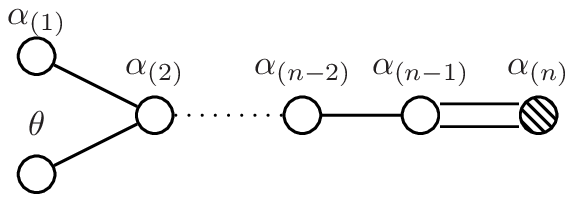}}
   \includegraphics{DynkinBnExtendedLabeled.eps}}\) \\
   & &  \\
   \(\mathrm{C}_n \) & \(  \mathfrak{sp}(2n)\) 
   & \(\CenterObject{\vphantom{\includegraphics[scale=1.1]
   {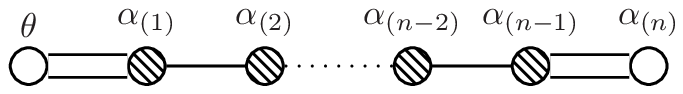}}
   \includegraphics{DynkinCnExtendedLabeled.eps}}\) \\
   & &  \\
   \(\mathrm{D}_n \) & \(  \mathfrak{so}(2n)\) 
   &
   \(\CenterObject{\includegraphics{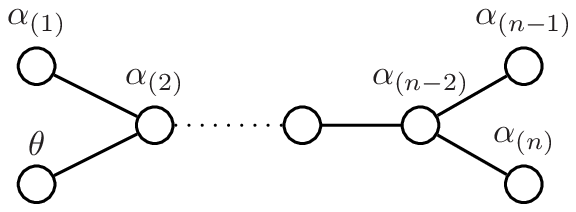}}\)\\
   \hline
  \end{tabular}
  \caption{The classical Lie algebras and the corresponding
        extended Dynkin diagrams. The shorter roots are hatched. If the simple 
        roots \(\SimpleRoot{i}\) and \(\SimpleRoot{j}\) enclose
        an angle of \(90^\circ\), \(120^\circ\) or \(135^\circ\),
        they are connected by 0, 1 or 2 lines, respectively.}
  \label{tab:DynkinDiagramsOfClassicalGroups}
 \end{center}
\end{table}

\begin{table}[ht*]
  \begin{center}
  \begin{tabular}{|c|l|}
   \hline 
   \multicolumn{1}{|c}{Name} & 
   \multicolumn{1}{|c|}{Extended Dynkin diagram} 
   \\
   \hline 
   & \\[-0.3mm]
   $\mathrm{G}_2$ &
   $\CenterObject{\includegraphics{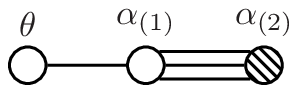}}$
   \\
   & \\
   $\mathrm{F}_4$ &
   $\CenterObject{\includegraphics{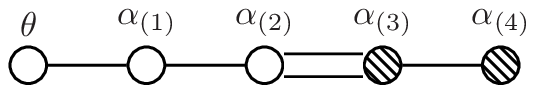}}$
   \\[1mm]
   & \\
   $\mathrm{E}_6$ &
   $\CenterObject{\includegraphics{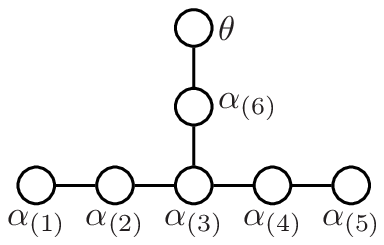}}$
   \\[1mm]
   & \\
   $\mathrm{E}_7$ &
   $\CenterObject{\includegraphics{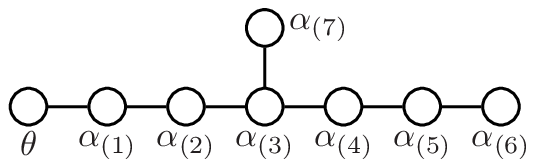}}$
   \\[1mm]
   & \\
   $\mathrm{E}_8$ &
   $\CenterObject{\includegraphics{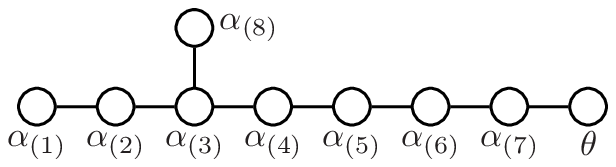}}$
   \\[1mm]
   & \\
   \hline
  \end{tabular}
 \end{center}
 \caption{The five exceptional Lie algebras. In \(\mathrm{G}_2\),
   the two simple roots enclose \(150^\circ\), which is indicated by 
   a triple line.}
 \label{tab:DynkinDiagramsOfExceptionalGroups}
\end{table}

\section{Obtaining all regular subgroups by orbifolding}\label{sg}

Orbifold GUTs~\cite{kaw,af,hn,hm,abc,hnos} are based on a gauge theory on 
$\mathbbm{R}^4\times M$, where $M$ is a manifold with some discrete 
symmetry group $K$. In addition to the action of $K$ on $M$, an action in 
internal space can be chosen using a homomorphism from $K$ to the automorphism
group of the Lie algebra of the gauge theory. If the classical field space 
is restricted by the requirement of $K$ invariance, a gauge theory on a
manifold with singularities, $M/K$, results in general. We assume that
$M/K$, though not necessarily $M$, is compact. At the singularities, which 
correspond to the fixed points of the space-time action of $K$, the gauge 
symmetry may be restricted (orbifold breaking). An early review of the 
structure of such models is contained in~\cite{hm1} (for more recent reviews 
see, e.g.,~\cite{Haba:2002py,orb}).

One of the main features of orbifold GUTs is the possibility of breaking 
a gauge group without the use of Higgs fields. The orbifold field theory
possesses the full (unified) gauge symmetry everywhere except for certain
fixed points. Although this fixed-point breaking is `hard', in the sense 
that the action does not possess the full gauge symmetry, gauge coupling 
unification is not lost due to the numerical dominance of the bulk. 
Furthermore, it is attractive for model building purposes 
that the symmetry  -- and hence the field content -- is characterized by 
different groups at different geometric locations, such as the various 
fixed points and the bulk. 

In this paper, we focus on inner-automorphism breaking, i.e., a 
homomorphism from $K$ to the gauge group $G$ together with the adjoint 
action of $G$ on itself is used to define the transformation of gauge 
fields under $K$.
Only gauge fields invariant under $K$ have zero modes. The corresponding 
generators define the symmetry of the low-energy effective theory, which is 
a subgroup of $G$. We will assume that \(G\) is simple since it is 
straightforward to extend our analysis to the product of simple groups and 
\(\mathrm{U}(1)\) factors.

To discuss the breaking in more detail, consider a group element $P$ which 
is the image of some element of $K$. Any $P\in G$ can be written as an 
exponential of some Lie algebra element and is therefore contained in some 
U(1) subgroup of $G$. Constructing a maximal torus starting from this U(1) 
and using the fact that the maximal torus in a compact Lie group is unique 
up to isomorphism~\cite{Fegan:1991jb}, it becomes clear that one can always 
write 
\begin{equation}
 P 
 \,=\, 
 \exp(-2\pi\I\, V\cdot \boldsymbol{H})\;,
\end{equation}
with some real vector $V$. Hence, the action of the gauge twist on the 
Lie algebra is given by
\begin{subequations}\label{eq:OnlyOneTwist}
\begin{eqnarray}
 P\,\boldsymbol{E}_\alpha\,P^{-1}
 & = & 
 \exp(-2\pi\I\,\alpha\cdot V)\,\boldsymbol{E}_\alpha
 \;,\label{eq:AbelianTwists4Ealpha}\\*
 P\,\boldsymbol{H}_i\,P^{-1}
 & = &
 \boldsymbol{H}_i\;.
\end{eqnarray}
\end{subequations}

We can also choose to write \(P=\exp(- 2\pi\I\,\xi\,\boldsymbol{T})\), 
where \(\boldsymbol{T}\) is a normalized Lie algebra element, 
\(\xi\in\mathbbm{R}\), and \(\xi\,\boldsymbol{T}=V\cdot \boldsymbol{H}\).
For generic $\xi$, $P$ commutes with precisely those Lie algebra elements 
with which $\boldsymbol{T}$ commutes. Thus, the breaking is the same as 
would follow from a Higgs VEV in the adjoint representation. 

However, it is clear from Eq.~(\ref{eq:AbelianTwists4Ealpha}) that, for
certain values of $\xi$, some of the \(\boldsymbol{E}_\alpha\) may pick 
up phases which are an integer multiple of \(2\pi\) and are thus left
invariant. In this case, the surviving subgroup is larger than the one
obtained from an adjoint VEV proportional to $\boldsymbol{T}$. This 
possibility is of particular interest since, in certain cases, such as the 
breaking of \(\mathrm{SO}(10)\) to 
\(\mathrm{SU}(4)\times\mathrm{SU}(2)\times\mathrm{SU}(2)\), the relevant 
subgroup can not be realized by using Higgs VEVs in the adjoint or any smaller 
representation.

\subsection{Orbifold-breaking to any maximal regular subgroup}
\label{o2ms}

We now show that, given a simple group \(G\) and a maximal regular\footnote{In 
this paper, we concentrate on the breaking to regular subgroups. For a 
discussion of non-regular embeddings (in the string theory context) see, 
e.g.,\cite{dmr}.}  subgroup \(H\), there exists a \(P\in G\) such that
\begin{equation}
 H \,=\, \{g\in G;\; P\,g\,P^{-1}=g\}\,.
\end{equation}
In other words, every maximal regular subgroup can be generated by an 
orbifold twist. 

In order to prove this statement, we first recall Dynkin's prescription
for generating semi-simple subgroups. It starts with the Dynkin diagram,
extends it by adding the most negative root, and then removes one of the 
simple roots, the resulting Dynkin diagram being that of a semi-simple 
subgroup. As demonstrated in~\cite{dyn3} (cf. Theorem 5.3), any 
maximal-rank, semi-simple subgroup of a given group can be obtained by 
successive application of this prescription. Maximal subgroups can always 
be obtained in the first application. 

To implement Dynkin's prescription and remove the simple root $\alpha_{(i)}$, 
one can use the fundamental weight $\mu^{(i)}$ and choose
\begin{equation}
 P \,=\, \exp\left(\frac{2\pi \I}{n}\,\frac{2}{|\SimpleRoot{i}|^2}\,
 \FundamentalWeight{i}\cdot\boldsymbol{H}
   \right)\;.
 \label{eq:FundamentalZnTwist}
\end{equation}
Obviously, $P$ commutes with all simple roots $\boldsymbol{E}_{\alpha_{(j)}}$ 
where $j\neq i$. 
To discuss the roots $\alpha_{(i)}$ and $\theta$, recall first that
\begin{equation}
 \theta = -\sum_{k=1}^r c_k\,\SimpleRoot{k}
\end{equation}
with the \(c_k\) being known as Coxeter labels. They can be read off from 
Tab.~\ref{tab:HighestWeightsOfAdjoints}. Such group-theoretical methods were
used in~\cite{kkkot} in the context of E$_8$ breaking in string theory. 

\begin{table}[!ht!]
 \begin{center}
  \begin{tabular}{|c|l|l|}
   \hline
   Group & Dynkin labels of \(\Lambda_\mathrm{ad}\) & Coxeter labels\\
   \hline
   \(\mathrm{A}_n\simeq\mathrm{SU}(n+1)\) & \((1,0,\dots 0,1)\) & 
   \((1,1,\dots 1)\)\\
   \(\mathrm{B}_n\simeq\mathrm{SO}(2n+1)\) & \((0,1,0,\dots)\) & 
   \((1,2,2,\dots 2,2)\)\\
   \(\mathrm{C}_n\simeq\mathrm{Sp}(2n)\) & \((2,0,0,\dots)\) & 
   \((2,2,\dots 2,1)\)\\
   \(\mathrm{D}_n\simeq\mathrm{SO}(2n)\) & \((0,1,0,\dots)\) & 
   \((1,2,2,\dots 2,1,1)\)\\
   \(\mathrm{G}_2\) & \((1,0)\) & \((2,3)\)\\
   \(\mathrm{F}_4\) & \((1,0,0,0)\) & \((2,3,4,2)\)\\
   \(\mathrm{E}_6\) & \((0,0,0,0,0,1)\) & \((1,2,3,2,1,2)\)\\
   \(\mathrm{E}_7\) & \((1,0,0,0,0,0,0)\) & \((2, 3, 4, 3, 2, 1, 2)\)\\
   \(\mathrm{E}_8\) & \((0,0,0,0,0,0,1,0)\) & \((2, 4, 6, 5, 4, 3, 2, 3)\)\\
   \hline
  \end{tabular}
 \end{center}
 \caption{Highest weights of the adjoint representations, 
  denoted by \(\Lambda_\mathrm{ad}\), 
  of the simple groups in the Dynkin basis and their
  Coxeter labels.}
 \label{tab:HighestWeightsOfAdjoints}
\end{table}

Thus, we can write the orbifold action on the two roots 
\(\boldsymbol{E}_{\SimpleRoot{i}}\) and \(\boldsymbol{E}_\theta\) as
\begin{subequations}
\begin{eqnarray}
 P\,\boldsymbol{E}_{\SimpleRoot{i}}\,P^{-1}
 & = &
 e^{ 2\pi\I/n}\,\boldsymbol{E}_{\SimpleRoot{i}}
 \;,\\*
 P\,\boldsymbol{E}_{\theta}\,P^{-1}
 & = &
 e^{- 2\pi\I\,c_i/n}\,\boldsymbol{E}_{\theta}
 \;,
\end{eqnarray}
\end{subequations}
which shows that, for \(\boldsymbol{E}_\theta\) to be invariant and 
\(\boldsymbol{E}_{\SimpleRoot{i}}\) to be projected out, we need \(c_i\ne 1\). 
Using Tab.~\ref{tab:HighestWeightsOfAdjoints} and the corresponding Dynkin
diagrams, it is easy to convince oneself that $c_i=1$ occurs only for 
those $i$ where the Dynkin-prescription with removal of $\alpha_{(i)}$ 
returns the original diagram. Thus, all non-trivial subgroups accessible by 
the Dynkin-prescription can be obtained by $\mathbbm{Z}_n$ orbifolding with 
$n=c_i$.

An interesting and subtle observation can be made in those cases where 
$c_i$ is not prime (only $c_i=4$ and $c_i=6$ occur). If $c_i=n=m\cdot k$,
a \(\mathbbm{Z}_m\) twist generated by $P^k$ is sufficient to project out 
$\boldsymbol{E}_{\SimpleRoot{i}}$ while keeping $\boldsymbol{E}_\theta$, 
yet the surviving subgroup is larger than for the corresponding 
\(\mathbbm{Z}_n\) twist \(P\) and its Dynkin diagram is not the one obtained 
by Dynkin's prescription. These are the famous five cases where Dynkin's 
prescription produces a subgroup that is not maximal~\cite{Golubitsky:1971ex}. 
They occur when removing the 3rd root of \(\mathrm{F}_4\), the 3rd root of 
\(\mathrm{E}_7\), and the 2nd, 3rd or 5th root of \(\mathrm{E}_8\).

It is easy to see that for $c_i$ prime the produced subgroup is maximal.
Indeed, the roots of $G$ which are not roots of the subgroup $H$ can be 
classified according to their `level' relative to $\alpha_{(i)}$, i.e.,
according to the coefficient of $\alpha_{(i)}$ in their decomposition 
in terms of simple roots. If a subgroup $H'$ with $H\subset H'\subset G$ 
exists, one of the levels below $c_i$ (which is the highest level) and 
above 1 must be occupied (i.e., its roots belong to $H'$). Let $\ell$ be the 
smallest of those levels. All multiples of $\ell$ are also occupied and, since 
$c_i$ is not a multiple, the difference between $c_i$ and one of those 
multiples must be smaller than $\ell$. However, by the way in which the 
commutation relations 
are realized in root space, the level corresponding to this difference must 
also be occupied. This is in contradiction to $\ell$ being the smallest 
occupied level in $H'$. 

Having dealt with all semi-simple maximal subgroups, we now come to maximal
subgroups containing $\mathrm{U}(1)$ factors. Given a maximal subgroup $H$ 
with $\mathrm{U}(1)$ factor, i.e., \(G\supset H\times\mathrm{U}(1)\), we 
can always break to a subgroup $H'$ by an adjoint VEV along this 
$\mathrm{U}(1)$ direction 
or a corresponding orbifold twist. It is obvious that $H\subset H'$ since, 
by the definition of $H$, all its elements commute with the generator of 
the above $\mathrm{U}(1)$. Thus, $H=H'$ and our analysis of orbifold breaking 
to all maximal-rank regular subgroups is complete. The maximal regular 
subgroups and the corresponding twists are listed in 
Tab.~\ref{tab:OrbifoldTwists} in App.~\ref{app:OrbifoldTwists}.
We would also like to mention that the maximal subgroups with 
\(\mathrm{U}(1)\) factors can be obtained by removing one node of the 
original Dynkin diagram which carries Coxeter label 1, and adding 
the \(\mathrm{U}(1)\) factor.

Now that it is clear how a given maximal regular subgroup can be generated 
by an orbifold twist, we can take the opposite point of view and ask to
which subgroups an arbitrary given gauge twist $P=\exp{(-2\pi\I\,\xi\,
\boldsymbol{T}})$ can lead. Since an adjoint VEV proportional to 
$\boldsymbol{T}$ breaks to a maximal rank subgroup $H\times$U(1), where the 
\(\mathrm{U}(1)\) is generated by $\boldsymbol{T}$, we can classify all 
$\boldsymbol{T}$'s by such maximal subgroups. These are given in various tables 
(see, in particular,~\cite{sla}) together with the branching rules for the 
adjoint representation
\begin{equation}
 \ad G 
 \,\to\, 
 \ad H \oplus \boldsymbol{1}(0) \oplus \boldsymbol{R}_1(q_1) \oplus 
 \boldsymbol{R}_2(q_2)
 \oplus\dots\;.
\end{equation}
Here the \(\boldsymbol{R}_i\) are representations under \(H\) and \(q_i\) 
the corresponding \(\mathrm{U}(1)\) charges. Under the gauge twist, 
the \(\boldsymbol{R}_i\) transform as \(\boldsymbol{R}_i\to e^{2\pi\I\,\xi 
q_i}\,\boldsymbol{R}_i\). This allows us to determine which particular sets 
of generators $\boldsymbol{R}_i$ survive for specific values of $\xi$, i.e.,
to identify those \(\boldsymbol{R}_i\) for which $\xi q_i=0\mod \mathbbm{Z}$. 
Together with the generators 
of $H\times$U(1), they form the Lie algebra of the new surviving subgroup
$H'\supset H\times$U(1). Thus, by analyzing all subgroups $H\times$U(1) and
all values of $\xi$, our classification is complete.

Finally, we would like to comment on the minimal order of the twist
required to achieve the breaking \(G\to H\). A very useful approximate
rule is that under a \(\mathbbm{Z}_n\) twist
\begin{equation}
 \dim H 
 \,\gtrsim\, 
 r + \frac{\dim G - r}{n}
 \;.
\end{equation}
The reason is that the \(r\) Cartan generators survive the twist anyway,
and the phases of the roots are proportional to a level relative to a
simple root \(\SimpleRoot{i}\), or linear combination of such levels.
Due to the symmetries of the root lattice, the phases are therefore 
almost evenly distributed among \(\{0,2\pi/n, 4\pi/n,\dots (n-1)2\pi/n\}\) 
where an excess at 0 is possible if the twist acts trivially on a certain
part of the algebra. An inspection of Tab.~\ref{tab:OrbifoldTwists} confirms 
our rule which becomes the more accurate the larger the group is.

\subsection{Some examples from the series $\boldsymbol{\mathrm{SO}(10)
\subset\mathrm{E}_6\subset\mathrm{E}_7\subset\mathrm{E}_8}$}

At this point, some examples are in order.
Let us start with the \(\mathrm{SO}(10)\) GUT which contains
the Georgi-Glashow group
\(G_\mathrm{GG}=\mathrm{SU}(5)\otimes\mathrm{U}(1)\)
and the Pati-Salam group
\(G_\mathrm{PS}=\mathrm{SU}(4)\times\mathrm{SU}(2)\times\mathrm{SU}(2)\)
as subgroups. These properties are nicely illustrated by using 
Dynkin's prescription:
\begin{figure}[!ht]
\begin{center}
 \subfigure[Extended Dynkin diagram of \(\mathrm{SO}(10)\).]
 {\begin{tabular}{c}
 \CenterObject{\includegraphics{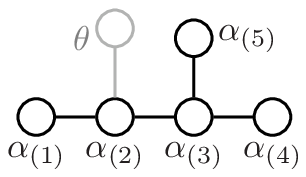}}\\
 {}
 \end{tabular}}
 \hfil
 \subfigure[Breaking to $G_\mathrm{PS}$.]
 {\begin{tabular}{c}
 \CenterObject{\includegraphics{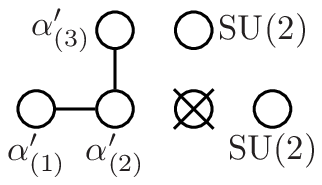}}\\
 {}
 \end{tabular}}
 \hfil
 \subfigure[Breaking to $G_\mathrm{GG}$.]
 {\begin{tabular}{c}
 \CenterObject{\includegraphics{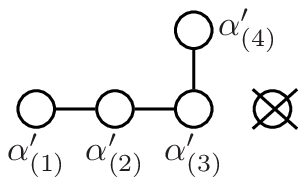}}\\
 {}
 \end{tabular}}
\end{center} 
 \caption{The breaking to the Pati-Salam and Georgi-Glashow subgroups 
  of \(\mathrm{SO}(10)\) can be illustrated by removing the 
  \(\SimpleRoot{3}\) (or \(\SimpleRoot{2}\)) node of the extended Dynkin 
  diagram (a) as shown in (b) or by removing \(\SimpleRoot{5}\) 
  (or \(\SimpleRoot{4}\)) as shown in (c).}
 \label{fig:ExtendedDynkinDiagramOfSO10}
\end{figure}
Starting from the extended Dynkin diagram 
(cf.~Fig.~\ref{fig:ExtendedDynkinDiagramOfSO10}), the diagram of
\(\mathrm{G}_\mathrm{PS}\) is obtained by deleting the third (or second)
node. Deleting the fourth (or fifth) node of the original diagram,
we arrive at \(G_\mathrm{GG}\). According to Sec.~\ref{o2ms}, twists
which break to \(G_\mathrm{GG}\) and \(\mathrm{G}_\mathrm{PS}\),
respectively, can be written as
\begin{subequations}\label{eq:PGGandPPS4SO10}
\begin{eqnarray}
 P_\mathrm{PS}
 & = &
 \exp(\pi\I\,\FundamentalWeight{3}\cdot\boldsymbol{H})
 \;,\\*
 P_\mathrm{GG}
 & = &
 \exp(\pi\I\,\FundamentalWeight{4}\cdot\boldsymbol{H})
 \;,
\end{eqnarray}
\end{subequations}
where we exploited the fact that \(|\SimpleRoot{i}|^2=2\) in
simply-laced groups. 

In \cite{abc,hnos}, it was shown that by identifying these two twists 
as generators of \(\mathbbm{Z}_2\times\mathbbm{Z}_2'\), the gauge symmetry 
on a \(\mathbbm{T}^2/(\mathbbm{Z}_2\times\mathbbm{Z}_2')\) orbifold is 
reduced to \(G_\mathrm{SM}'=\mathrm{SU}(3)\times\mathrm{SU}(2)\times
\mathrm{U}(1)_\mathrm{Y}\times\mathrm{U}(1)_\chi\subset\mathrm{SO}
(10)\). The resulting geometry can be visualized as a `pillow' with the 
corners corresponding to the fixed points.

The relevant group theory can be understood as follows: 
\(\FundamentalWeight{3}\cdot\boldsymbol{H}\) and 
\(\FundamentalWeight{4}\cdot\boldsymbol{H}\) are the \(\mathrm{U}(1)\) 
generators appearing in \(G_\mathrm{SM}'\). The corresponding decomposition 
of the adjoint representation of \(\mathrm{SO}(10)\) reads
\begin{eqnarray}
 \boldsymbol{45}
 & \to &
 (\boldsymbol{1},\boldsymbol{1})_{(6,4)}
 \oplus(\overline{\boldsymbol{3}},\boldsymbol{1})_{(-4,4)}
 \oplus(\boldsymbol{3},\boldsymbol{2})_{(1,4)}
 \oplus(\boldsymbol{3},\boldsymbol{2})_{(-5,0)}
 \nonumber\\*
 & & {}
 \oplus(\boldsymbol{1},\boldsymbol{1})_{(-6,-4)}
 \oplus(\boldsymbol{3},\boldsymbol{1})_{(4,-4)}
 \oplus(\overline{\boldsymbol{3}},\boldsymbol{2})_{(-1,-4)}
 \oplus(\overline{\boldsymbol{3}},\boldsymbol{2})_{(5,0)}
 \nonumber\\
 & & {}
 \oplus(\boldsymbol{8},\boldsymbol{1})_{(0,0)}
 \oplus(\boldsymbol{1},\boldsymbol{3})_{(0,0)}
 \oplus(\boldsymbol{1},\boldsymbol{1})_{(0,0)}
 \oplus(\boldsymbol{1},\boldsymbol{1})_{(0,0)}
 \;,\label{eq:DecompositionOf45}
\end{eqnarray}
where the \(\mathrm{SU}(3)\times\mathrm{SU}(2)\) representations are given 
in boldface and the U(1) charges \((q_Y,q_\chi)\) appear as index.
The twist which is responsible for this breaking is generated
by a linear combination of the generators of the two \(\mathrm{U}(1)\)s,
and rotates the charged representations by a phase \(2\pi\,(y\,q_Y + 
\chi\,q_\chi)\). For some combinations of \(\chi\) and \(y\), the orbifold 
breaking preserves a larger symmetry than adjoint breaking. For example, 
if \((\boldsymbol{1},\boldsymbol{1})_{(6,4)}\), \((\boldsymbol{1}, 
\boldsymbol{1})_{(-6,-4)}\), \((\overline{\boldsymbol{3}}, 
\boldsymbol{1})_{(-4,4)}\) and \((\boldsymbol{3},\boldsymbol{1})_{(4,-4)}\) 
survive (e.g., by taking \(\chi=0\) and \(y=1/2\)), the resulting gauge group 
is \(G_\mathrm{PS}\). If, on the other hand, 
\((\boldsymbol{3},\boldsymbol{2})_{(-5,0)}\) and 
\((\overline{\boldsymbol{3}},\boldsymbol{2})_{(5,0)}\) survive (e.g., by 
taking \(\chi=1/8\) and \(y=0\)), the resulting gauge group is 
\(G_\mathrm{GG}\). It is then clear that \(G_\mathrm{SM}'\) results as 
an intersection of gauge fields surviving \(P_\mathrm{PS}\) and 
\(P_\mathrm{GG}\). The breaking to \(G_\mathrm{SM}'\) can also
be realized on a single \(\mathbbm{Z}_4\) fixed point, e.g., by
using \(\chi=1/16\) and \(y=1/4\).

As a side-remark, let us restate the above discussion in terms of matrices:
Consider the adjoint VEV 
\begin{equation}\label{eq:GeorgisVev}
 v
 \,=\,
 \diag(a,a,a,b,b)\otimes
 \left(\begin{array}{cc}0 & 1 \\ -1 & 0\end{array}\right)
 \;,
\end{equation}
which breaks \(\mathrm{SO}(10)\) to \(G_\mathrm{SM}'\) \cite{Georgi:1999jb}.
For the special case \(a=\pm b\), the remaining symmetry is larger and equal 
to \(G_\mathrm{GG}\). Alternatively, these breakings can be realized by 
a gauge twist \(P=\exp[2\pi\I\,v]\) at an orbifold fixed point. In this 
case, taking \(a=0\) and \(b=1/2\) yields \(P_\mathrm{PS}\) and \(a=b=1/4\)
yields \(P_\mathrm{GG}\).

Let us now turn to the task of extending the `pillow' of Asaka, 
Buchm\"uller and Covi~\cite{abc1} along the chain of exceptional groups
\(\mathrm{SO}(10)\subset\mathrm{E}_6\subset\mathrm{E}_7\subset\mathrm{E}_8\).
A related discussion has already appeared in \cite{Haba:2002vc}. However,
as will become clear below, we disagree with some of the results of that
paper. 

The obvious generalizations of
\eqref{eq:PGGandPPS4SO10} for the exceptional groups read 
\begin{subequations}
\begin{eqnarray}
 P_\mathrm{GG}^{(r)}
 & := &
 \exp(\pi\I\,\FundamentalWeight{r-1}\cdot\boldsymbol{H})
 \;,\\*
 P_\mathrm{PS}^{(r)}
 & := &
 \exp(\pi\I\,\FundamentalWeight{r-2}\cdot\boldsymbol{H})
 \;.
\end{eqnarray}
\end{subequations}
In other words, the generalizations of \(P_\mathrm{GG}\) and \(P_\mathrm{PS}\)
to higher groups along the above chain remove the nodes \(\SimpleRoot{r-1}\) 
and \(\SimpleRoot{r-2}\) respectively. This is illustrated in 
Tab.~\ref{tab:GGPS4Es}.
\begin{table}[!ht]
 \[
  \begin{array}{|l|l|}
   \hline
    \CenterObject{\includegraphics{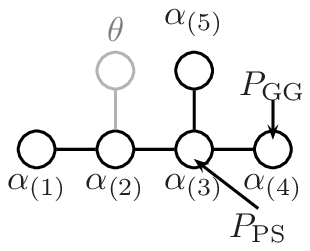}}
        &
        \begin{array}{l}
        \begin{array}{rcl}
         \mathrm{SO}(10) 
         & \xrightarrow{P_\mathrm{GG}\phantom{\big|}} & 
         \mathrm{SU}(5)\times\mathrm{U}(1)\\
         \mathrm{SO}(10) 
         & \xrightarrow{P_\mathrm{PS}} & 
         \mathrm{SU}(4)\times\mathrm{SU}(2)\times\mathrm{SU}(2)\\
         \mathrm{SO}(10) 
         & \xrightarrow{P_\mathrm{GG}\cdot P_\mathrm{PS}} & 
         \mathrm{SU}(5)'\times\mathrm{U}(1)'
                 \end{array}\\
         G_\mathrm{GG}\cap G_\mathrm{PS}
         \, =  \,
         \mathrm{SU}(3)\times\mathrm{SU}(2)\times\mathrm{U}(1)^2
         \phantom{\Big|}
        \end{array}
        \\
   \hline
    \CenterObject{\includegraphics{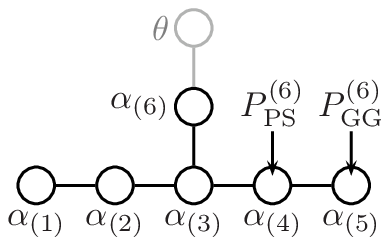}}
        &
                \begin{array}{l}
        \begin{array}{rcl}
         \mathrm{E}_6
         & \xrightarrow{P_\mathrm{GG}^{(6)}\phantom{\big|}} & 
         \mathrm{SO}(10)\times\mathrm{U}(1)\\
         \mathrm{E}_6
         & \xrightarrow{P_\mathrm{PS}^{(6)}} & 
         \mathrm{SU}(6)\times\mathrm{SU}(2)\\
         \mathrm{E}_6 
         & \xrightarrow{P_\mathrm{GG}^{(6)}\cdot P_\mathrm{PS}^{(6)}} & 
         \mathrm{SO}(10)'\times\mathrm{U}(1)'
                 \end{array}\\
         G_\mathrm{GG}^{(6)}\cap G_\mathrm{PS}^{(6)}
         \,= \,
         \mathrm{SU}(5)\times\mathrm{U}(1)^2 \phantom{\Big|}
        \end{array}
        \\   
   \hline
    \CenterObject{\includegraphics{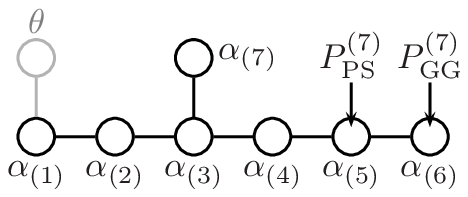}}
        &
                \begin{array}{l}
        \begin{array}{rcl}
         \mathrm{E}_7
         & \xrightarrow{P_\mathrm{GG}^{(7)}\phantom{\big|}} & 
         \mathrm{E}_6\times\mathrm{U}(1)\\
         \mathrm{E}_7
         & \xrightarrow{P_\mathrm{PS}^{(7)}} & 
         \mathrm{SO}(12)\times\mathrm{SU}(2)\\
         \mathrm{E}_7 
         & \xrightarrow{P_\mathrm{GG}^{(7)}\cdot P_\mathrm{PS}^{(7)}} & 
         \mathrm{E}_6'\times\mathrm{U}(1)'
                 \end{array}\\
         G_\mathrm{GG}^{(7)}\cap G_\mathrm{PS}^{(7)}
         \, = \,
         \mathrm{SO}(10)\times\mathrm{U}(1)^2 \phantom{\Big|}
        \end{array}
        \\   
   \hline
   \CenterObject{\includegraphics{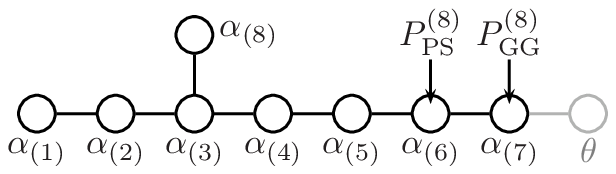}}
        &
                \begin{array}{l}
        \begin{array}{rcl}
         \mathrm{E}_8
         & \xrightarrow{P_\mathrm{GG}^{(8)}\phantom{\big|}} & 
         \mathrm{E}_7\times\mathrm{SU}(2)\\
         \mathrm{E}_8
         & \xrightarrow{P_\mathrm{PS}^{(8)}} & 
         \mathrm{E}_7'\times\mathrm{SU}(2)'\\
         \mathrm{E}_8 
         & \xrightarrow{P_\mathrm{GG}^{(8)}\cdot P_\mathrm{PS}^{(8)}} & 
         \mathrm{E}_7^{\prime\prime}\times\mathrm{SU}(2)^{\prime\prime}
                 \end{array}\\
         G_\mathrm{GG}^{(8)}\cap G_\mathrm{PS}^{(8)}
         \, = \,
         \mathrm{E}_6\times\mathrm{U}(1)^2 \phantom{\Big|}
        \end{array}
        \\   
   \hline
  \end{array}
 \]
 \caption{Breaking patterns to the Georgi-Glashow and Pati-Salam like 
 subgroups in \(\mathrm{SO}(10)\) and the three exceptional groups 
 \(\mathrm{E}_6\), \(\mathrm{E}_7\) and \(\mathrm{E}_8\).}
 \label{tab:GGPS4Es}
\end{table}

The breaking patterns of \(\mathrm{SO}(10)\), \(\mathrm{E}_6\) and
\(\mathrm{E}_7\) are easily determined by the use of Dynkin's prescription 
because the Coxeter label corresponding to the nodes removed by 
\(P_\mathrm{GG}^{(r)}\) and \(P_\mathrm{PS}^{(r)}\) are 1 and 2, respectively.
In the case of \(\mathrm{E}_8\), it is also easy to see that  
\(P_\mathrm{GG}^{(8)}\) breaks to \(\mathrm{E}_7\times\mathrm{SU}(2)\). 
For \(P_\mathrm{PS}^{(8)}\), the pattern is not so obvious: Since the
6th Coxeter label is 3 (cf.\ Tab.~\ref{tab:HighestWeightsOfAdjoints}),
and we use a \(\mathbbm{Z}_2\) twist, the second level in terms of
\(\SimpleRoot{6}\) survives \(P_\mathrm{PS}^{(8)}\), but
\(\theta\) is projected out. We see that the subgroup must contain
\(\mathrm{E}_6\) and \(\mathrm{SU}(2)\), and it can not be
\(\mathrm{E}_6\times\mathrm{SU}(3)\) because this is not a symmetric 
subgroup of \(\mathrm{E}_8\). Hence, it must be also 
\(\mathrm{E}_7\times\mathrm{SU}(2)\).\footnote{This 
is in contradiction to the breaking pattern given 
in~\cite{Haba:2002vc}. Assigning negative parity
to \((\boldsymbol{27},\boldsymbol{3})\) and 
\((\overline{\boldsymbol{27}},\overline{\boldsymbol{3}})\) is inconsistent
since, as can be seen from the commutator 
\([(\boldsymbol{27},\boldsymbol{3}),(\boldsymbol{27},\boldsymbol{3})]
\subset (\overline{\boldsymbol{27}},\overline{\boldsymbol{3}})\),
it does not correspond to an algebra automorphism. This commutator does
not vanish since \((\boldsymbol{27},\boldsymbol{3})\oplus
(\overline{\boldsymbol{27}},\overline{\boldsymbol{3}})\) contains two
positive levels with respect to \(\SimpleRoot{6}\), linked
by the raising operator \(\boldsymbol{E}_{\SimpleRoot{6}}\).} 
We checked this statement by using a computer algebra system.
In \cite{hnos}, an interesting property of the \(\mathrm{SO}(10)\)
twists was pointed out: \(P_\mathrm{GG}'=P_\mathrm{GG}\cdot P_\mathrm{PS}\) 
breaks to a different \(\mathrm{SU}(5)\times\mathrm{U}(1)\) subgroup of 
\(\mathrm{SO}(10)\), where the simple factor is often called `flipped 
\(\mathrm{SU}(5)\)'. This property is maintained for all three exceptional 
groups:  \(P_\mathrm{GG}^{(r)}\cdot P_\mathrm{PS}^{(r)}=P_\mathrm{GG}^{(r)\,
\prime}\). Here \(P_\mathrm{GG}^{(r)\,\prime}\) breaks to a subgroup linked 
by an inner automorphism to the subgroup left invariant by 
\(P_\mathrm{GG}^{(r)}\). The reason is that \(P_\mathrm{GG}^{(r)\,\prime}=
\exp[\pi\I\,(\FundamentalWeight{r-1}
+\FundamentalWeight{r-2})\cdot\boldsymbol{H}]\) commutes with 
\(\boldsymbol{E}_{\SimpleRoot{r-1}+\SimpleRoot{r-2}}\) which then 
becomes a simple root of the subgroup, and projects out \(\boldsymbol{E}_{\theta}\). 
This root encloses an angle of \(120^\circ\) with \(\SimpleRoot{r-3}\) 
so that the resulting Dynkin diagram coincides with the one obtained
by employing \(P_\mathrm{GG}\). In the simple root system arising from
the substitution \((\SimpleRoot{r-2},\SimpleRoot{r-1})\to
(\SimpleRoot{r-2}+\SimpleRoot{r-1},-\SimpleRoot{r-1})\),
\(P_\mathrm{GG}^{(r)\,\prime}\) acts in the same way as 
\(P_\mathrm{GG}^{(r)}\) in the original root system.

\section{Rank reduction and non-Abelian twists}\label{rr}

It is obvious from the discussion so far that using only one
inner-automorphism orbifold twist can never result in rank reduction.
We therefore investigate the possibilities which arise when
two (or more) twists are applied. Rank reduction of the gauge group was 
proposed in the context of string theory in \cite{Ibanez:1987xa}.
Here, we will discuss this issue in the context of field theory,
where one has fewer group-theoretic and geometric constraints. 

We assume that we have an additional orbifold symmetry,
\begin{equation}\label{eq:NonAbelianTwist}
 P' 
 \,=\, 
 \exp \left[-2\pi\I \,\xi\,(\boldsymbol{E}_\beta+\boldsymbol{E}_{-\beta})
 \right]
 \quad\text{or}\quad
 P' 
 \,=\, 
 \exp \left[2\pi\,\xi \,(\boldsymbol{E}_\beta-\boldsymbol{E}_{-\beta})\right]
 \;,
\end{equation}
where \(\I(\boldsymbol{E}_\beta+\boldsymbol{E}_{-\beta})\) and 
\(\boldsymbol{E}_\beta-\boldsymbol{E}_{-\beta}\) are real generators 
outside the Cartan subalgebra. For simplicity, let us focus on the case 
that \(\beta\) is a simple root, i.e.,
\begin{equation}
 P'_j
 \,=\,
 \exp \left[-2\pi\I \,\xi\,
        (\boldsymbol{E}_{\SimpleRoot{j}}+\boldsymbol{E}_{-\SimpleRoot{j}})
 \right]
 \;.
\end{equation}
Then the raising and lowering operators 
\(\boldsymbol{E}_{\pm\SimpleRoot{j}}\) form an \(\mathrm{SU}(2)\)
group together with 
\(\boldsymbol{h}=\SimpleRoot{j}\cdot \boldsymbol{H}\).
Clearly, this linear combination of Cartan generators
`rotates' under the action of \(P'\) like
\begin{equation}
 P_j^{\prime\,-1}\,\boldsymbol{h}\,P_j'
 \,=\,
 \cos(4\pi\,\xi)\,\boldsymbol{h}
 -\I\,\sin(4\pi\xi)\,(\boldsymbol{E}_{\SimpleRoot{j}}-\boldsymbol{E}_{-\SimpleRoot{j}})
 \;,\label{eq:TransformationOfCartanGenerators}
\end{equation}
where we restricted ourselves to the case that \(\SimpleRoot{j}\) has length 
\(\sqrt{2}\).
Since a linear combination of Cartan generators transforms non-trivially,
it is obvious that rank reduction is possible.
Note also that these rotations yield an extension of the well-known
Weyl reflections, i.e., the reflections with respect to a plane perpendicular
to a simple root.

It is straightforward, but somewhat tedious to derive the
action on arbitrary roots \(\boldsymbol{E}_\alpha\).
In simply-laced gauge groups, the root chains have at most length two unless 
they contain Cartan generators. Thus, \(N_{\alpha,\pm\SimpleRoot{j}}\ne 0\) 
implies \(N_{\alpha,\mp\SimpleRoot{j}}=0\). For the upper sign, we obtain 
(for \(\alpha\ne\SimpleRoot{i}\))
\begin{equation}
 P^{\prime \,-1}_j\,\boldsymbol{E}_\alpha\,P'_j  
 \,=  \,
 \cos(2\pi\,N_{\alpha,\SimpleRoot{j}}\,\xi)
        \,\boldsymbol{E}_\alpha
 +\I\sin(2\pi\,N_{\alpha,\SimpleRoot{j}}\,\xi)\,
        \boldsymbol{E}_{\alpha+\SimpleRoot{j}}
 \;,    
\end{equation}
where we use the normalization constants \(N_{\alpha,\beta}\) 
as defined in equation \eqref{eq:DefOfNalphabeta} with the convention to
choose them positive.

{}From the discussion so far, it is clear that we can break any simple group
factor completely by non-Abelian twists: The roots can always be removed
by suitable exponentials of the Cartan generators, and the \(\boldsymbol{H}_i\)
can be projected out by using Eq.~\eqref{eq:TransformationOfCartanGenerators}.
This observation has an obvious application: Let $H\subset G$ 
be the subgroup that we want to obtain by orbifolding. Let $H'\subset G$ be
the maximal subgroup that commutes with $H$ and the Cartan generators of 
which are orthogonal to the Cartan generators of $H$. If $H'$ is semi-simple, 
an orbifold breaking to $H$ is always possible. In this context, it is 
interesting to observe that \(\mathrm{E}_8\) is the only simple group 
containing a maximal regular subgroup of the form \(\mathrm{SU}(5)\times 
H'\) with \(H'\) semi-simple, namely $\mathrm{E}_8\supset\mathrm{SU}(5)\times
\mathrm{SU}(5)$. Thus, one could say that \(\mathrm{E}_8\) is 
the smallest GUT group larger than \(\mathrm{SU}(5)\) which can be 
orbifolded to the SM without additional \(\mathrm{U}(1)\) factors.

A further example is in order: It is clear that we can break an 
\(\mathrm{SU}(2)\) factor completely by the methods described above. Thus,
since \(\mathrm{E}_6\supset\mathrm{SU}(6)\times\mathrm{SU}(2)\),
we can achieve \(\mathrm{E}_6\to\mathrm{SU}(6)\)
by taking \(P=\exp[\pi\I\,\SimpleRoot{1}\cdot\boldsymbol{H}]\)
and \(P'=\exp[\pi\I\,(\boldsymbol{E}_{\SimpleRoot{1}}+
\boldsymbol{E}_{-\SimpleRoot{1}})/2]\).
In addition, we can modify \(P\) in a way so that the breaking is stronger,
e.g., \(\mathrm{E}_6\to\mathrm{SU}(5)\times\mathrm{U}(1)\).

However, if extra \(\mathrm{U}(1)\) factors are contained in $H'$,
the story becomes more complicated. One is tempted to conclude that such 
extra factors can not be removed, given that this is obviously not possible 
by adjoint VEV breaking. However, in the case of orbifold breaking 
this is not true. Consider, for example, \(\mathrm{SO}(5)\)
which can be broken to \(\mathrm{SU}(2)\times\mathrm{U}(1)\)
by using \(P=\diag(-1,-1,1,1,1)\). The extra \(\mathrm{U}(1)\) can be
destroyed by invoking \(P'=\diag(1,-1,-1,-1,-1)\).
This example is particularly interesting since here \(P\) and \(P'\) commute
although the rank is reduced (which is possible because the corresponding 
generators 
do not commute). The above SO(5) example is special because \(P'\), which 
maps the U(1) generator to minus itself, acts on the other (real) 
representations in a way consistent with SU(2) symmetry. If we deal with 
complex representations, i.e., the adjoint of \(G\) branches as
\begin{equation}
 \ad G 
 \, \to \, 
 \ad H \oplus\mathbbm{1}(0)\oplus\boldsymbol{R}(q)\oplus
\overline{\boldsymbol{R}}(-q)
 \oplus \dots\;,\label{cr}
\end{equation}
where \(\boldsymbol{R}(q)\) and \(\overline{\boldsymbol{R}}(-q)\) are 
conjugate to each other, a flip of the \(\mathrm{U}(1)\) charge carries 
\(\boldsymbol{R}\) and \(\overline{\boldsymbol{R}}\) into each other.
The flip then acts non-trivially on \(H\) so that flipping the
\(\mathrm{U}(1)\) factor without affecting \(H\) is impossible. 

We emphasize that this excludes the possibility of 
orbifold breaking of the U(1) factor in a large class of cases. Namely, 
let $H\times$U(1)$\subset G$ such that the U(1) is the maximal group 
commuting with $H$. Clearly, any automorphism of $G$ leaving $H$ invariant 
has to map the U(1) onto itself. Since the only non-trivial automorphism 
of U(1) is the above sign flip, the presence of complex representations
$H$ in the adjoint of $G$ (cf.~Eq.~\ref{cr}) excludes the required 
$H$-preserving autmorphism of $G$. The extension to $H\times H'\subset 
G$, where $H'$ is a product group containing U(1) factors, is 
straightforward. 

\begin{figure}[!h]
 \begin{center}
 \subfigure[$\mathbbm{T}^2/D_2$\label{fig:D2Torus}]
        {\CenterObject{\includegraphics{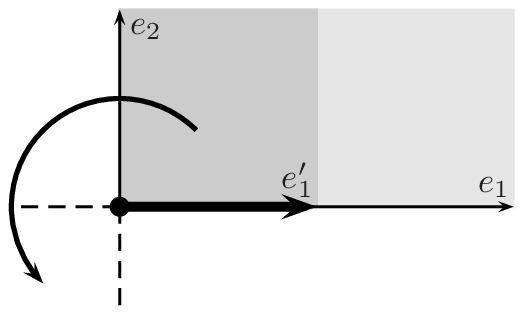}}}
 \hfil
 \subfigure[$\mathbbm{T}^2/D_3$\label{fig:D3Torus}]
        {\CenterObject{\includegraphics{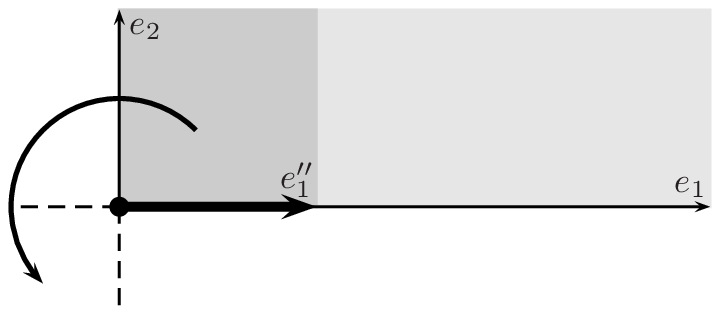}}}
 \end{center}
 \caption{Examples of  (a) a \(\mathbbm{T}^2/D_2\) and (b) a 
 $\mathbbm{T}^2/D_3$
 orbifold where rank reduction is possible. The action of the
 dihedral group consists in a rotation by \(180^\circ\) around the origin, 
 and in a translation by \(e_1'\) in case (a), and a translation by 
 \(e_1^{\prime\prime}\) or \(2\,e_1^{\prime\prime}\) in case (b).}
 \label{fig:DihedralExamples}
\end{figure}

The above \(\mathrm{SO}(5)\) scenario with \(P\) and \(P'\) can, for
example, be realized in \(4+2\) dimensions with compact space 
\(\mathbbm{T}^2/(\mathbbm{Z}_2\times\mathbbm{Z}_2')\). The \(\mathbbm{Z}_2\) 
generator acts on the torus as a rotation by \(180^\circ\), the 
\(\mathbbm{Z}_2'\) generator acts as a shift by half of one of the original 
torus translations (cf.~Fig.~\ref{fig:D2Torus}).

It turns out that the elements of \(\mathbbm{Z}_2\times\mathbbm{Z}_2'\)
comply with the multiplication law of the dihedral group $D_2$ of order 
4.\footnote{Recall that the dihedral group of order \(2n\), called \(D_n\), 
can be envisaged as the group generated by the rotation of a regular 
\(n\)-polygon by 
\(2\pi/n\) and the flip over one of its edges \cite{Wigner}. Clearly,
the dihedral group always can be embedded in an 
\(\mathrm{SO}(3)\simeq\mathrm{SU}(2)\).
Anomalies of dihedral orbifolds are discussed in \cite{GrootNibbelink:2003gd}.}
While the dihedral group of order
4 is Abelian, higher order dihedral groups are not. We illustrate a 
possible way of using the order 6 group $D_3$ in an orbifold construction in 
Fig.~\ref{fig:D3Torus}. It follows the 
\(\mathbbm{T}^2/(\mathbbm{Z}_2\times\mathbbm{Z}_2')\) construction up to
the fact that we now divide the cell into three parts instead of two.
Embedding it into a gauge group then allows for realizing non-Abelian twists.

These examples can be generalized in the following way: The orbifold can be 
interpreted as \(\mathbbm{O}=\mathbbm{T}^n/R\) where \(R\) is a symmetry
of the lattice, and the torus arises by modding out flat space by
discrete translations, \(\mathbbm{T}^n=\mathbbm{R}^n/\Lambda\). 
By embedding the full symmetry group \(K\), containing the operations of 
\(R\) as well as the translations, into the gauge group, it is possible 
to achieve that the torus \(\mathbbm{T}^n\), which arises as 
intermediate step in this picture, carries Wilson lines \cite{Ibanez:1986tp}.
Since the generators associated with the Wilson lines do not necessarily
commute with the twists corresponding to embedding the operations of \(R\)  
into the gauge group, rank reduction is possible \cite{Ibanez:1987xa}.
We believe that similar constructions will be important for model building.

Let us briefly comment on non-regular embeddings. Consider the group
\(\mathrm{SU}(3)\) which contains \(\mathrm{SO}(3)\) (the subgroup of real 
matrices) as an S-subgroup (in Dynkin's terminology). Let us pick two 
generators of the embedded \(\mathrm{SO}(3)\), for instance
\begin{equation}
 T_1
 \,=\,
 \left(\begin{array}{ccc}
        0 & 1 & 0\\
        -1 & 0 & 0 \\
        0 & 0 & 0
 \end{array}\right)
 \quad\text{and}\quad
 T_2
 \,=\,
 \left(\begin{array}{ccc}
        0 & 0 & -1\\
        0 & 0 & 0 \\
        1 & 0 & 0
 \end{array}\right)
 \;.
\end{equation}
It is then straightforward to convince oneself that imposing the twists
\(P_1=\exp(2\pi\I\,T_1/4)\) and \(P_2=\exp(2\pi\I\,T_2/4)\) breaks
\(\mathrm{SU}(3)\) completely. Similar constructions can be used to
break larger groups with only a few twists. For instance, 
\(\mathrm{E}_8\) has a maximal S-subalgebra \(\mathfrak{su}(2)\) and
can therefore be broken completely by only two twists, e.g.,
by embedding a suitable dihedral group in the \(\mathrm{SU}(2)\).

\section{Conifold GUTs}
\label{coni}

We now want to continue the discussion of the generic structure of orbifold 
GUTs given at the beginning of Sec.~\ref{sg} and show that a mild 
generalization of the construction principles leads to a much larger 
freedom in model building. Our main focus will be on 6d models.

\subsection{Geometry and gauge symmetry breaking}

In 5 dimensions, the geometry is very constraining. Up to isomorphism, the 
only smooth compact manifold is $\mathbbm{S}^1$, where one has the familiar 
problems of obtaining chiral matter and of fixing the Wilson line, the value 
of which represents a modulus which, in the SUSY setting, can not be stabilized 
by perturbative effects. The only compact orbifold is the interval, which can 
always be viewed as $\mathbbm{S}^1/(\mathbbm{Z}_2\times \mathbbm{Z}_2')$ (with 
$\mathbbm{S}^1/\mathbbm{Z}_2$ being a special case). The gauge breaking at 
each boundary is determined by a $\mathbbm{Z}_2$ automorphism and can be 
interpreted as explicit breaking by boundary conditions. One may try to 
generalize the setting by considering breaking by a boundary localized 
Higgs (in the limit where the VEV becomes large)~\cite{nsw} or ascribing 
Dirichlet and Neumann boundary conditions to different gauge fields (without 
the $\mathbbm{Z}_2$ automorphism restriction)~\cite{hm1}. Furthermore, it 
is possible to ascribe the breaking to a singular Wilson line crossing the 
boundary~\cite{hm1}. However, it appears to be unavoidable that geometry is 
used only in a fairly trivial way and that the breaking is confined entirely 
to the small-scale physics near the brane, outside the validity range of 
effective field theory.

Group-theoretically, the 5d setting is also fairly constrained since the 
relative orientation of the gauge twists at the two boundaries is a 
modulus. To be more specific, let $P_1=\exp({\boldsymbol{T}_1})$ and 
$P_2=\exp({\boldsymbol{T}_2})$ be the two relevant twists. Even though this 
makes rank reduction possible in principle, we are faced with the problem 
that, if the Wilson line connecting the boundaries develops an appropriate 
VEV, the situation becomes equivalent to both $\boldsymbol{T}_1$ and 
$\boldsymbol{T}_2$ being in the Cartan subalgebra, in which case the 
symmetry is enhanced to a maximal-rank subgroup. SUSY prevents 
the modulus from being fixed by loop corrections.\footnote{For more 
details and a discussion of the non-supersymmetric case see~\cite{heb} 
and~\cite{Haba:2002py} respectively.}

\begin{figure}[!ht]
 \begin{center}
 \CenterObject{\includegraphics{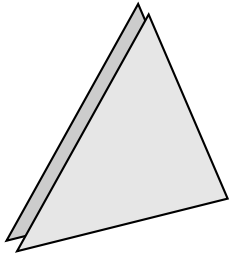}}
 \end{center}
 \caption{Construction of a compact manifold with singularities from two 
triangles.}\label{sm}
\end{figure}

In 6 dimensions, the situation is much more complicated and interesting. 
Clearly, the smooth torus has the 
same problems as the $\mathbbm{S}^1$ discussed above. However, there is a
large number of
compact manifolds with conical singularities. A simple way to envisage such 
singular manifolds or, more precisely, conifolds is given in Fig.~\ref{sm}.
The fundamental space consists of two identical triangles. The geometry 
is determined by gluing together the edges of the depicted triangles, 
thus leading to a triangle with a front and a back, a triangular 
`pillow'. It is flat everywhere except for the three conical singularities 
corresponding to the three corners of the basic triangle. Each deficit 
angle is $2(\pi-\varphi)$, where $\varphi$ is the corresponding angle of the 
triangle. Obviously, in this construction the basic triangle can be replaced 
by any polygon. If the polygon is non-convex, negative deficit angles 
appear.

Four specific polygons deserve a separate discussion. These are the 
rectangle, the equilateral triangle, the isosceles triangle with a 
$90^{\circ}$ angle, and the triangle with angles $30^\circ$, $60^\circ$, and 
$90^\circ$. The conifolds constructed in the above manner from these 
polygons can alternatively be derived from the torus as a $\mathbbm{Z}_2$, 
$\mathbbm{Z}_3$, $\mathbbm{Z}_4$ and $\mathbbm{Z}_6$ orbifold, respectively.
Given that $\mathbbm{Z}_n$ can not be a symmetry 
of a 2-dimensional lattice for $n>6$, it is clear that this last method of 
constructing conifolds is highly constrained when compared to the generic 
conifold of Fig.~\ref{sm} with an arbitrary polygon. However, from the 
perspective of effective field theory model building, there appears to 
be no fundamental reason to discard the multitude of possibilities arising 
in the more general framework. 

Clearly, even more possibilities open up if, in addition to conical 
singularities one allows for 1-dimensional boundaries. These arise in 
orbifolding if a $\mathbbm{Z}_2$ reflection symmetry (in contrast to the 
$\mathbbm{Z}_n$ rotation symmetries above) of the torus is modded 
out~\cite{hnos} (see also~\cite{li}). However, in what follows we will 
concentrate on construction with conical singularities only. 

We now turn to the possibilities of geometric gauge symmetry breaking 
on conifolds. Recall first that, if a given conifold can be constructed 
from a smooth manifold by modding out a discrete symmetry group, i.e.,
as an orbifold, then 
an appropriate embedding of this discrete group into the automorphism 
group of the gauge Lie algebra will lead to a gauge 
symmetry reduction. Working directly on the fundamental space (as opposed to 
the covering space) this gauge breaking can be ascribed to non-trivial 
values of Wilson lines encircling each of the conical singularities. 

It is now fairly obvious how to introduce this type of breaking in the 
generic construction of Fig.~\ref{sm} (possibly with the triangle replaced 
by an arbitrary polygon). First, we identify one edge of the front polygon
with the corresponding edge of the back polygon. 
Next, when identifying along the two adjacent edges,
one uses the freedom of introducing a relative gauge twist $P\in G$. In more 
detail, if $(x,y)$ and $(x',y')$ parametrize front and back polygon near
the relevant edge (such that the edge is at $y=0$ or $y'=0$), one demands
$A'(x'=x,y'=0)=P\,A(x,y=0)\,P^{-1}$ for the gauge potentials $A$ and $A'$ on 
the two polygons. Continuing with the identifications, one finds that there 
is a freedom of choosing $n-1$ gauge twists $P_i$ in the presence of $n$ 
conical singularities. Technically, this is due to the fact that the 
identification along one of the edges can always be made trivial using 
global gauge rotations of one of the polygons. A geometric understanding 
follows from the fact that the global topology is that of a sphere, in which 
case the Wilson lines around $n-1$ singularities fix the last Wilson line
(we always assume the vacuum configuration, i.e., $A$ is locally pure gauge). 

Clearly, we want to obtain a smooth manifold (except for the singularities) 
in the end so that, to be more precise, the $P_i$ have to be introduced in 
the appropriate transition functions of the defining atlas. However, we 
believe that it is not necessary to spell out this familiar construction 
in detail. 

Instead of using only inner automorphisms described by $P_i$, we could have 
allowed for outer automorphisms in the transition functions. In this case, 
which we will not pursue in this paper, the corresponding vacua are clearly 
disconnected from those defined only by inner automorphisms. The theory can 
then be thought of as defined on a generalization of a principal bundle (in the 
commonly used definition of principal bundles the transition functions involve 
only inner automorphisms).

We now want to analyze the gauge fields in a small open subset including one 
conical singularity. A convenient parametrization is given by polar 
coordinates $(r,\varphi)$ with $0<r<\epsilon$ and $0\leq\varphi<\beta$, where the 
singularity is at $r=0$ and the deficit angle is $2\pi-\beta$. As familiar 
from the Hosotani mechanism on smooth manifolds~\cite{hos}, we can trade the 
gauge twist in the matching from $\varphi=\beta$ to $\varphi=0$ for a background 
gauge field which, for a twist $P=\exp(\boldsymbol{T})$, can be chosen as
$A=\boldsymbol{e}_\varphi\boldsymbol{T}/(\beta r)$. Here $\boldsymbol{e}_\varphi$ 
is the unit-vector in $\partial_\varphi$ direction so that $A$ is a 
Lie-algebra-valued vector. This simple exercise demonstrates explicitly 
that, at least locally, the breaking can be attributed to a non-vanishing 
gauge field VEV in a flat direction. However, in contrast to the Hosotani 
mechanism, the corresponding modulus can be fixed without violating the 
locality assumption (which we consider as fairly fundamental in effective 
field theory). Namely, the value of the Wilson line described by the 
above $A$ can be determined by some unspecified small-distance physics
directly at the singularity. This is similar to the boundary breaking 
in 5 dimensions. In contrast to the 5d case, however, the breaking at the 
conical singularity is visible to the bulk observer, who can encircle the 
singularity and measure the Wilson line without coming close to the 
singularity. Thus, one might be tempted to conclude that this type of 
breaking has a better definition in terms of low-energy effective field 
theory. 

To conclude this subsection, we want to collect the generalizations of 
6d field theoretic orbifold models discussed above. First, one can work 
on conifolds, i.e., use deficit angles that can not result from modding out 
on the basis of a smooth manifold. The gauge twist at each singularity may,
however, be still required to be consistent with the geometric twist. 
Second, one can insist on conventional orbifolds as far as the geometry 
is concerned but use arbitrary gauge twists at each conical singularity,
i.e., give up the connection between the rotation angles in tangent and 
gauge space. Third, one may drop both constraints and work on conifolds 
with arbitrary deficit angles and gauge twists. Obviously, such 
constructions can also be carried out in more than 6 dimensions. 
The detailed discussion of those is beyond the scope of the present paper.

\subsection{Generating chiral matter}\label{chir}

In general, compactification on a non-flat manifold can provide 
chiral matter if the holonomy group of the compact manifold fulfills
certain criteria. For example, it is well-known that compactification
of a 10d SYM theory on Calabi-Yau manifolds with \(\mathrm{SU}(3)\) 
holonomy~\cite{Candelas:en} or on orbifolds~\cite{dhvw} leads to 
\(N=1\) SUSY in 4d.
Both constructions are not unrelated as many orbifolds can be regarded
as singular limits of manifolds in which the curvature is concentrated at the
fixed points. Since the reduction of SUSY is a matter of geometry,
compactification of a higher-dimensional field theory on a conifold
can also lead to \(N=1\) supersymmetric models in 4d.

Interesting models have been constructed using the fact that
the vector multiplet of \(N=(1,1)\) SUSY in 6d corresponds to
one vector and three chiral multiplets in 4d language,
\(A=(V,\phi_1,\phi_2,\phi_3)\). The fact that three copies of chiral 
multiplets appear automatically may be an explanation of the observed 
number of generations \cite{wy}. The above 6d theory can be interpreted as 
arising from a 10d SYM, in which case the scalars of the chiral 
multiplets are the extra components of gauge fields~\cite{mss}, for example, 
\(\phi_1\ni A_5+\I A_6\), \(\phi_2\ni A_7+\I A_8\) and \(\phi_3\ni 
A_9+\I A_{10}\) (\(A_M\) denote the components of the 10d vector).
When defining our 6d models, we require the field transformations 
associated with going around a conical singularity to be an element 
of an SU(3) subgroup of the full $\mathrm{SO}(6)\simeq\mathrm{SU}(4)$ 
symmetry of the 
underlying 10d SYM theory. Under this subgroup, which we call \(\mathrm{SU}(3)\) 
R-symmetry, the chiral superfields $\phi_i$ transform as a 
$\boldsymbol{3}$.\footnote{For more details see, e.g.,~\cite{mss} 
as well as~\cite{wy,Imamura:2001es}.}
The appealing feature of such a construction is that matter 
multiplets are not put in `by hand' but arise in a natural way from a 
higher-dimensional SYM theory \cite{Babu:2002ti,wy,gmn,nb}.

The action of the \(\mathrm{SU}(3)\) R-symmetry transformation on the chiral 
superfields is not completely arbitrary. For example, if \(\phi_1\ni A_5+\I A_6\), 
the transformation of \(\phi_1\) is fixed by geometry, e.g.,
when modding out a rotation symmetry, a corresponding rotation has to be 
applied to the \(\phi_1\) superfield. Thus, when going around a conical 
singularity, \(\phi_1\) receives a phase which is given by \(2\varphi\), 
where \(\varphi\) is the corresponding angle of the polygon.
Since multiplying by the phase \(e^{\I 2\varphi}\) corresponds to a rotation
in the complex plane, we will call \(2\varphi\) `rotation angle' in what
follows. Clearly, the rotation angle sums up with the deficit angle to 
\(2\pi\).

\section{Specific models}\label{mo}

Let us now discuss three models in which some of the main features of the 
last sections are exemplified. All these models are based on a SYM 
theory in \(4+2\) dimensions endowed with \((N_1,N_2)=(1,1)\) SUSY.
In 4d we then deal with three chiral superfields
\(\phi_1\), \(\phi_2\) and \(\phi_3\) where we assume that
\(\phi_1=A_5+\I A_6\) so that the action of the R-symmetry on \(\phi_1\) is
fixed (cf.\ Sec.~\ref{chir}).

\subsection{$\boldsymbol{\mathrm{E}_7\to\mathrm{SU}(5)\times\mathrm{SU}(3)_\mathrm{F}\times\mathrm{U}(1)}$}

Consider a SYM theory based on an \(\mathrm{E}_7\) gauge group.
\(\mathrm{E}_7\) contains 
\(\mathrm{SU}(5)\times\mathrm{SU}(3)_\mathrm{F}\times\mathrm{U}(1)\),
and the adjoint representation decomposes as
\begin{eqnarray} 
 \boldsymbol{133}
 & \to &
 (\boldsymbol{24},\boldsymbol{1})_0
 \oplus
 (\boldsymbol{1},\boldsymbol{1})_{0}
 \oplus
 (\boldsymbol{1},\boldsymbol{8})_{0}
 \oplus
 (\boldsymbol{5},\boldsymbol{1})_6
 \oplus
 (\overline{\boldsymbol{5}},\boldsymbol{1})_{-6}
 \nonumber\\*
 & & {}
 \oplus
 (\boldsymbol{10},\overline{\boldsymbol{3}})_{-2}
 \oplus
 (\overline{\boldsymbol{5}},\boldsymbol{3})_{-4}
 \oplus
 (\overline{\boldsymbol{10}},\boldsymbol{3})_{2}
 \oplus
 (\boldsymbol{5},\overline{\boldsymbol{3}})_{4}
 \;,\label{eq:DecompositionOfE7}
\end{eqnarray}
where we use a notation analogous to Eq.~\eqref{eq:DecompositionOf45}.

As explained in Sec.~\ref{sg}, the twist \(P\) which causes the desired breaking
can be understood as exponential of the \(\mathrm{U}(1)\) generator.
Under this twist, the multiplets appearing in Eq.~\eqref{eq:DecompositionOfE7}
acquire phases which are proportional to the \(\mathrm{U}(1)\) charge. 
By taking the proportionality constant to be \(-1/12\), we arrive at 
the phases listed in Tab.~\ref{tab:PhasesE7toSU5} where here and below phases 
are given in units of \(2\pi\).\footnote{The twist can be thought of as 
\(P=\diag(\omega,\omega,\omega,\omega,\omega,-\omega)\),
with \(\omega\) defined as the 12th root of 1,
acting on the \(\mathrm{SU}(6)\) embedded in \(\mathrm{E}_7\).
Although the action of \(P\) on a fundamental representation
of \(\mathrm{SU}(6)\) would be the one of a \(\mathbbm{Z}_{12}\) twist,
its action on \(\mathrm{E}_7\) is \(\mathbbm{Z}_6\) since the adjoint
of \(\mathrm{E}_7\) only contains antisymmetric and adjoint representations
of \(\mathrm{SU}(6)\).}

\begin{table}[!ht]
 \begin{center}
  $\begin{array}{|rcl|rcl|rcl|}
   \hline
    (\boldsymbol{24},\boldsymbol{1})_{0} & : & \boldsymbol{0}
        &
        (\boldsymbol{5},\boldsymbol{1})_{6} & : & \boldsymbol{1/2}
        &
        (\overline{\boldsymbol{5}},\boldsymbol{3})_{-4} & : & \boldsymbol{1/3}
                \vphantom{\overline{\overline{\boldsymbol{5}}}}
        \\
        (\boldsymbol{1},\boldsymbol{1})_{0} & : & \boldsymbol{0}
        & 
        (\overline{\boldsymbol{5}},\boldsymbol{1})_{-6} & : & \boldsymbol{1/2}
        & 
        (\overline{\boldsymbol{10}},\boldsymbol{3})_{2} & : & 5/6
        \\
        (\boldsymbol{1},\boldsymbol{8})_{0} & : & \boldsymbol{0}
        &
        (\boldsymbol{10},\overline{\boldsymbol{3}})_{-2} & : & \boldsymbol{1/6}
        &
        (\boldsymbol{5},\overline{\boldsymbol{3}})_{4} & : & 2/3\\
   \hline
  \end{array}$
 \end{center}
 \caption{Phases (in units of \(2\pi\)) for the different multiplets of
 \(\mathrm{SU}(5)\times\mathrm{SU}(3)_\mathrm{F}\times\mathrm{U}(1)\subset\mathrm{E}_7\).
 Zeros correspond to the surviving gauge bosons, other phases which are
 compensated by the R-symmetry transformations are written in boldface.}
 \label{tab:PhasesE7toSU5}
\end{table}

The smallest phase present is \(1/6\) so that \(P\) is a \(\mathbbm{Z}_6\) 
twist. Therefore, the R-symmetry acts on \(\phi_1\) as a \(-60^\circ\)
rotation in the 5-6 plane, and thus the 
\((\boldsymbol{10},\overline{\boldsymbol{3}})_{-2}\) possesses a zero-mode. 
We choose the transformation of \(\phi_2\) such that the
\((\overline{\boldsymbol{5}},\boldsymbol{3})_{-4}\) survives as well,
and the phase of \(\phi_3\) is then fixed by the determinant condition.
More explicitly, by taking
\begin{equation}
 R
 \,=\,
 \exp[2\pi\I\,\diag(-1/6,2/3,-1/2)]
 \;\in\mathrm{SU}(3)
 \;,
\end{equation}
we can achieve that 3 generations of \(\boldsymbol{10}\) and
\(\overline{\boldsymbol{5}}\) survive without any mirrors, indicated
by boldface phases in Tab.~\ref{tab:PhasesE7toSU5},
and an \(N=1\) SUSY in 4d is preserved.
It is also interesting to observe that the only additional
surviving superfields, namely \((\boldsymbol{5},\boldsymbol{1})_6\) and 
\((\overline{\boldsymbol{5}},\boldsymbol{1})_{-6}\) which acquire phases
\(1/2\) and therefore have zero-modes due to the third diagonal entry 
of \(R\), carry the quantum numbers of the light Higgs fields
in the supersymmetric \(\mathrm{SU}(5)\) theory. Thus, the \(\mathrm{SU}(5)\)
part of this model looks relevant for reality, and is in particular
anomaly-free. 

\begin{figure}[!ht]
\begin{center}
 \CenterObject{\includegraphics{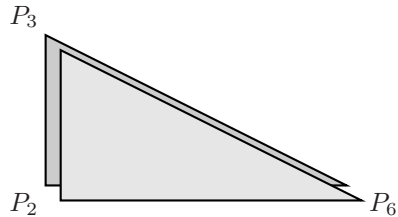}}
\end{center}
 \caption{Example of a \(4+2\) dimensional orbifold allowing for
 \(\mathbbm{Z}_6\) twists. The fundamental space consists of two triangles. 
 The geometry can be illustrated by gluing together the edges of the 
 depicted triangles, thus leading to a triangle with a front- and a backside.}
 \label{fig:DihedralTriangle}
\end{figure}

The geometry of this model, which can be constructed as a standard orbifold 
\(\mathbbm{T}^2/\mathbbm{Z}_6\), is given by two triangles with
angles \(30^\circ\), \(60^\circ\) and \(90^\circ\) 
(cf.\ Fig.~\ref{fig:DihedralTriangle}). The \(\mathbbm{Z}_6\) twist 
\(P\) (\(P_6=P\) in Fig.~\ref{fig:DihedralTriangle}) is associated with
the first of these fixed points; the twists \(P^2\) and \(P^3\) are 
associated with the remaining two fixed points. By construction, the order 
of rotation in the two extra dimensions coincides with the order of the 
twist in the gauge group.
It is then straightforward to determine the gauge groups which survive
at these fixed points. In the actual example, they turn out to be
\(\mathrm{SU}(6)\times\mathrm{SU}(3)_\mathrm{F}\) and \(\mathrm{SU}(8)\),
respectively. The content of non-vanishing fields at these fixed points 
is also found to be anomaly-free under the relevant surviving gauge group 
in both cases, which implies the absence of localized anomalies~\cite{abca}. 

The complete model is, however, not free of anomalies. This is due to 
localized anomalies at the $P_6$ fixed points, where the gauge group is 
SU(5)$\times$U(1)$\times\mathrm{SU}(3)_\mathrm{F}$. However, the 
SU(5) part by itself is free of localized anomalies even at this fixed 
point. Thus, if \(\mathrm{SU}(3)_\mathrm{F}\times\mathrm{U}(1)\) is
broken, as it has to be in order to describe reality, there are no 
anomalies. The desired breaking of the unwanted symmetries may be due
to fields which live on the fixed points, however, discussing such
possibilities is beyond the scope of this study. 
Note also that if we were to break the additional symmetry by rank-reducing
twists, fewer matter fields would survive. That is also the reason why
we do not use rank-reducing twists in the next two models.

\subsection{$\boldsymbol{\mathrm{E}_8\to G_\mathrm{PS}\times \mathrm{U}(1)^3}$}

Under \(\mathrm{E}_8\to\mathrm{SO}(10)\times\mathrm{SU}(4)\), the 
adjoint representation of \(\mathrm{E}_8\) decomposes like    
\begin{equation}
 \boldsymbol{248}
 \,\to\,
 (\boldsymbol{45},\boldsymbol{1})\oplus(\boldsymbol{1},\boldsymbol{15})
 \oplus(\boldsymbol{16},\boldsymbol{4})
 \oplus(\overline{\boldsymbol{16}},\overline{\boldsymbol{4}})
 \oplus(\boldsymbol{10},\boldsymbol{6})
 \;.
 \label{eq:248toGPSxSU4}
\end{equation}  
\(\mathrm{SO}(10)\) contains the Pati-Salam group \cite{ps}
\(G_\mathrm{PS}=\mathrm{SU}(4)\times\mathrm{SU}(2)\times\mathrm{SU}(2)\)
whereby
\begin{subequations}
\begin{eqnarray}
 \boldsymbol{45}
 & \to & 
 (\boldsymbol{15},\boldsymbol{1},\boldsymbol{1})\oplus
 (\boldsymbol{1},\boldsymbol{3},\boldsymbol{1})\oplus
 (\boldsymbol{1},\boldsymbol{1},\boldsymbol{3})\oplus
 (\boldsymbol{6},\boldsymbol{2},\boldsymbol{2})
 \;,\\*
 \boldsymbol{16}
 & \to &
 (\boldsymbol{4},\boldsymbol{2},\boldsymbol{1})\oplus
 (\overline{\boldsymbol{4}},\boldsymbol{1},\boldsymbol{2})
 \;,\\
 \boldsymbol{10}
 & \to &
 (\boldsymbol{6},\boldsymbol{1},\boldsymbol{1})\oplus
 (\boldsymbol{1},\boldsymbol{2},\boldsymbol{2})
 \;.
\end{eqnarray}
\end{subequations}
This breaking can be achieved by using the rotation 
\(-\mathbbm{1}\in\mathrm{SU}(2)\) for the first (or the second) 
\(\mathrm{SU}(2)\). In addition, we can now impose the twist
\begin{equation}
 F
 \,=\,
 \exp \left[2\pi\,\I\,
        \diag\left(1/3,-1/6,x,-1/6-x\right)
 \right]
 \in\mathrm{SU}(4)
 \;,
\end{equation}
where, e.g., \(x=\frac{1}{4}\),
in order to break \(\mathrm{SU}(4)\to[\mathrm{U}(1)]^3\).
The charges of the \(\boldsymbol{4}\) are then given by 
\(q_i\in\{1/3,-1/6,x,-1/6-x\}\). The charges of the \(\boldsymbol{6}\) 
are \(q_i+q_j\) with \(i\ne j\) since the \(\boldsymbol{6}\) is the 
antisymmetric part of the \(\boldsymbol{4}\times\boldsymbol{4}\) of 
\(\mathrm{SU}(4)\), and finally the charges of the \(\boldsymbol{15}\)
are \(q_i-q_j\).

Together with the R-symmetry transformation
\begin{equation}
 R
 \,=\,
 \exp\left[2\pi\I\,\diag\left(-1/3,-1/3,-1/3\right)\right]
 \;,
\end{equation}
three chiral generations of matter and three Higgs, i.e.,
\((\boldsymbol{1},\boldsymbol{2},\boldsymbol{2})\), survive. 
Only for certain \(x\), additional fields will possess zero-modes, and
we will choose \(x\) to equal none of these values.
Here, the number of generations is due to dimensional reduction
of \((N_1,N_2)=(1,1)\) SUSY in 6d to 4d.
The surviving gauge group is 
\(G_\mathrm{PS}\times[\mathrm{U}(1)]^3\).
The geometry is given by an equilateral triangle with
the three corners corresponding to three identical fixed points.

Obviously, for such a construction, the geometric twist, i.e., the
rotation in the two extra dimensions, is of a lower order than the
group theoretical twist. This requires going beyond the usual 
field-theoretic orbifold constructions (although the geometry is still 
an orbifold).
As proposed in Sec.~\ref{coni}, we define a field theory on a manifold
with three conical singularities, each of them possessing a deficit 
angle of \(2\pi/3\). This construction is then an equilateral
triangle. We then add Wilson lines such that the group-theoretical
twist \(P\) at two of the fixed points equals the one described above.
The twist at the third fixed point is then constrained to be $P^{-2}$ by 
the global geometry. 

At each singularity of the conifold, the Pati-Salam part of the gauge 
group is anomaly-free. This is obvious for first two fixed points since 
the non-vanishing fields are those of the standard model with three Higgs
doublets. At the third fixed point, the gauge symmetry is enhanced to the 
group SO(10), which has no 4d anomalies.\footnote{
Quite 
generally, the anomaly at a given conical singularity can be 
calculated from the zero-mode anomaly by considering a conifold where this 
specific singularity appears several times (possibly together with other 
conical singularities, the anomalies of which are already 
known)~\cite{ggnow}. However, we do not investigate this further in the 
present paper. For recent work on the explicit calculation of anomalies in 
6d models see~\cite{abca,gq,GrootNibbelink:2003gd}.
}
Again, investigating mechanisms to break the extra \(\mathrm{U}(1)\)s as well
as \(G_\mathrm{PS}\) to \(G_\mathrm{SM}\) is beyond the scope of this paper.

Note finally that this particular model can be viewed as an extension 
of~\cite{wy}, where three generations arise from the three chiral 
superfields present in the 4d description of a 10d SYM theory, i.e., they
follow from the presence of three complex extra dimensions.\footnote{
It 
has been claimed that this is related to the mechanism for obtaining three 
generations used in the string theory models reviewed in~\cite{far}.
}
The new points in 
our construction are the doublet-triplet splitting solution arising from the 
breaking to the Pati-Salam group (see~\cite{dm} and the recent related 
stringy models of~\cite{kim1}) and the realization of all rather than just 
part of the matter fields in terms of the SYM multiplet.

\subsection{$\boldsymbol{\mathrm{E}_8\to G_\mathrm{PS}\times
\mathrm{SU}(3)_\mathrm{F}\times\mathrm{U}(1)}$}

Alternatively, we can obtain \(G_\mathrm{PS}\) from \(\mathrm{E}_8\) 
and maintain an \(\mathrm{SU}(3)_\mathrm{F}\) flavour symmetry by
breaking the extra \(\mathrm{SU}(4)\) of the
decomposition \eqref{eq:248toGPSxSU4} to 
\(\mathrm{SU}(3)_\mathrm{F}\times\mathrm{U}(1)\).
In order to achieve this breaking, we take a central element 
of \(\mathrm{SU}(3)\),
\begin{equation}
 P
 \,=\,
 \exp\left[2\pi\I\,\left(1/3,1/3,1/3,0\right)\right]
 \;.
\end{equation}
The phases which arise by combining this twist with 
\(\exp(- 2\pi\I/4\,\mathbbm{1})\in\mathrm{SU}(4)\) are listed 
in Tab.~\ref{tab:PhasesE8toGPSxSU3F}.
\begin{table}[!ht]
 \[
 \begin{array}{|rcl|rcl|rcl|}
 \hline
 & & & & & & & & \\[-0.3cm]
 (\boldsymbol{15},\boldsymbol{1},\boldsymbol{1};\boldsymbol{1}_0)
 & : & \boldsymbol{0}
 &
 (\boldsymbol{1},\boldsymbol{1},\boldsymbol{1};\overline{\boldsymbol{3}}_{-4/3})
 & : & 2/3
 &
 (\overline{\boldsymbol{4}},\boldsymbol{1},\boldsymbol{2}; \overline{\boldsymbol{3}}_{-1})
 & : & 11/12
 \\
 (\boldsymbol{1},\boldsymbol{3},\boldsymbol{1};\boldsymbol{1}_0)
 & : & \boldsymbol{0}
 &
 (\boldsymbol{4},\boldsymbol{2},\boldsymbol{1};\boldsymbol{3}_{1})
 & : & \boldsymbol{1/12}
 &
  (\overline{\boldsymbol{4}},\boldsymbol{1},\boldsymbol{2};\boldsymbol{1}_{-3})
 & : & 1/4 
 \\
 (\boldsymbol{1},\boldsymbol{1},\boldsymbol{3};\boldsymbol{1}_0)
 & : & \boldsymbol{0}
 &
 (\boldsymbol{4},\boldsymbol{2},\boldsymbol{1};\boldsymbol{1}_{-3})
 & : & 3/4 
 &
 (\boldsymbol{6},\boldsymbol{1},\boldsymbol{1};\boldsymbol{3}_{1})
 & : & 5/6
 \\
 (\boldsymbol{1},\boldsymbol{1},\boldsymbol{1};\boldsymbol{8}_0)
 & : & \boldsymbol{0}
 &
 (\overline{\boldsymbol{4}},\boldsymbol{1},\boldsymbol{2};\boldsymbol{3}_{1})
 & : &  \boldsymbol{7/12}
 &
 (\boldsymbol{6},\boldsymbol{1},\boldsymbol{1};\overline{\boldsymbol{3}}_{-1})
 & : & 1/6 
 \\
 (\boldsymbol{1},\boldsymbol{1},\boldsymbol{1};\boldsymbol{1}_0)
 & : & \boldsymbol{0}
 &
 (\overline{\boldsymbol{4}},\boldsymbol{1},\boldsymbol{2};\boldsymbol{1}_{3})
 & : & 1/4
 & 
 (\boldsymbol{1},\boldsymbol{2},\boldsymbol{2};\boldsymbol{3}_{1})
 & : & \boldsymbol{1/3}
 \\
 (\boldsymbol{6},\boldsymbol{2},\boldsymbol{2};\boldsymbol{1}_0)
 & : & 1/2
 &
 (\boldsymbol{4},\boldsymbol{2},\boldsymbol{1};\overline{\boldsymbol{3}}_{-1})
 & : &  11/12 
 &
 (\boldsymbol{1},\boldsymbol{2},\boldsymbol{2};\overline{\boldsymbol{3}}_{-1})
 & : & 2/3
 \\
 (\boldsymbol{1},\boldsymbol{1},\boldsymbol{1};\boldsymbol{3}_{4/3})
 & : & \boldsymbol{1/3} 
 &  
 (\boldsymbol{4},\boldsymbol{2},\boldsymbol{1};\boldsymbol{1}_{3})
 & : &  3/4
 & & &\\
 \hline
\end{array} 
\]
\caption{Table of the phase factors for the different multiplets
 of \(G_\mathrm{PS}\times\mathrm{SU}(3)_\mathrm{F}\times\mathrm{U}(1)\subset 
 \mathrm{E}_8\). Zeros correspond to surviving gauge bosons; other phases
 which are compensated by the R-symmetry transformation are written
 in boldface.
 }
\label{tab:PhasesE8toGPSxSU3F}
\end{table}
Now let us simultaneously impose an R-symmetry twist
\begin{equation}
 R\,=\,
 \exp\left[2\pi\I\,\diag(-1/12,-7/12,-1/3)\right]
 \;.
\end{equation}
It is then easy to see from Tab.~\ref{tab:PhasesE8toGPSxSU3F} that the
zero modes which emerge in the matter sector are three generations 
of SM matter, three Higgs and three additional neutrinos.

In order to realize such a model, we have again to relax the constraints
of usual orbifold models, and therefore consider a manifold with a conical
singularity with deficit angle \(2\pi\cdot 5/12\) instead (cf.\ Sec.~\ref{coni}).
To be more specific, we envisage the geometry of the model as an isosceles
triangle with an angle of \(2\pi\cdot 5/12\).
Each corner corresponds to a fixed point,
and we are free to choose both \(\pi/12\) fixed points identically.
By construction, the group-theoretical twist \(P\) at the \(\pi/12\) fixed
points generate a \(\mathbbm{Z}_{12}\), i.e., \(P^{12}=\mathbbm{1}\).
At the remaining \(2\pi\cdot 5/12\) `corner', we choose the twist 
\(P^{10}=P^{-2}\) for consistency. Interestingly, a quick inspection of 
Tab.~\ref{tab:PhasesE8toGPSxSU3F} reveals that the there surviving
gauge symmetry is \(\mathrm{SO}(10)\). Obviously, the \(\mathrm{SO}(10)\)
part of the gauge theory at this fixed point is anomaly-free automatically.

Once more, discussing the breaking of the extra gauge symmetry is beyond 
the scope of this study.

\section{Conclusions}\label{co}

We have explored some of the group-theoretical possibilities in orbifold GUTs.
In particular, we showed that, given a simple gauge group \(G\), 
the breaking to any maximal-rank regular subgroup can be achieved by 
orbifolding. 

We further studied rank reduction and found that
simple group factors can always be broken completely.
This is possible when using non-Abelian twists, and also if twists
commute but the corresponding generators do not.
Using such constructions in orbifolding is made possible by embedding
a non-Abelian (or even Abelian) space group into the gauge group.

We then extended the familiar concept of orbifold GUTs by replacing the
orbifolds by manifolds with conical singularities. The possibilities we 
discussed include orbifold geometries endowed with unrestricted Wilson 
lines wrapping the conical singularities, manifolds with conical 
singularities with arbitrary deficit angles, and combinations thereof.

Finally, we presented three specific models where three generations of fields
carrying the SM quantum numbers come from a SYM theory in 6d. While the
first one is a conventional orbifold model illustrating the usefulness of 
our group theoretical methods, the two others are based on the two new 
concepts mentioned above. 

To summarize, we explored several new and interesting methods and 
possibilities which can be used in orbifold GUTs and their generalizations.

As none of our models is yet completely realistic, more effort is required 
in order to discuss phenomenological consequences. However, it is very 
appealing how easily three generations can be obtained and the 
doublet-triplet splitting problem can be solved. Thus, promoting our
models to realistic ones in future studies appears to be worthwhile.

\noindent {\bf Note added:} While this paper was being finalized, 
Ref.~\cite{kim} 
appeared where Dynkin diagram techniques were used as well. Aspects of our 
analysis not addressed by~\cite{kim} include, in particular, the breaking 
of any simple group to all maximal-rank regular subgroups, rank-reduction, 
as well as several new field-theoretic concepts and models.

\vspace*{-.3cm}
\section*{Acknowledgments}

\vspace*{-.3cm}
We would like to thank Fabian Bachmaier, Wilfried Buchm\"{u}ller, John 
March-Russell, Hans-Peter Nilles, Mathias de Riese and Marco Serone for 
useful discussions.

\renewcommand{\thesection}{\Alph{section}}
\renewcommand{\thesubsection}{\Alph{section}.\arabic{subsection}}

\def\theequation{\Alph{section}.\arabic{equation}}

\renewcommand{\thetable}{\Alph{section}.\arabic{table}}
\setcounter{section}{0}

\vspace*{-.3cm}
\section{Table of orbifold twists}\label{app:OrbifoldTwists}

\vspace*{-.3cm}
\begin{table}[!ht]
\begin{center}
 \begin{tabular}{|l|c|l|l|}
  \hline
  Group & Twist & Symmetric subgroup & Comment\\
  \hline
  \hline
   \(\mathrm{SU}(N+M)\) & \(\mathbbm{Z}_2\) & 
        \(\mathrm{SU}(N)\times\mathrm{SU}(M)\times\mathrm{U}(1)\) & \\
  \hline
  \(\mathrm{SO}(N+M)\) & \(\mathbbm{Z}_2\) & 
   \(\mathrm{SO}(N)\times\mathrm{SO}(M)\) & \(N\) or \(M\) even\\
  \(\mathrm{SO}(2N)\) & \(\mathbbm{Z}_2\) & 
   \(\mathrm{SU}(N)\times\mathrm{U}(1)\) & \\
  \hline
  \(\mathrm{Sp}(2N+2M)\) & \(\mathbbm{Z}_2\) & 
        \(\mathrm{Sp}(2N)\times\mathrm{Sp}(2M)\) & \\
  \(\mathrm{Sp}(2N)\) & \(\mathbbm{Z}_2\) & 
        \(\mathrm{SU}(N)\times\mathrm{U}(1)\) & \\
  \hline
  \hline
  \(\mathrm{G}_2\) & \(\mathbbm{Z}_2\) & 
        \(\mathrm{SU}(2)\times\mathrm{SU}(2)\) & \\
  \(\mathrm{G}_2\) & \(\mathbbm{Z}_3\) & 
        \(\mathrm{SU}(3)\) & \\ 
  \hline
  \(\mathrm{F}_4\) & \(\mathbbm{Z}_2\) & 
        \(\mathrm{Sp}(6)\times\mathrm{SU}(2)\) & \\
  \(\mathrm{F}_4\) & \(\mathbbm{Z}_3\) & 
        \(\mathrm{SU}(3)\times\mathrm{SU}(3)\) & \\
  \(\mathrm{F}_4\) & \(\mathbbm{Z}_4\) & 
        \(\mathrm{SU}(4)\times\mathrm{SU}(2)\) & not maximal\\
  \(\mathrm{F}_4\) & \(\mathbbm{Z}_2\) & 
        \(\mathrm{SO}(9)\) & \\
  \hline
  \(\mathrm{E}_6\) & \(\mathbbm{Z}_2\) & 
        \(\mathrm{SO}(10)\times\mathrm{U}(1)\) & \\
  \(\mathrm{E}_6\) & \(\mathbbm{Z}_2\) & 
        \(\mathrm{SU}(6)\times\mathrm{SU}(2)\)& \\
  \(\mathrm{E}_6\) & \(\mathbbm{Z}_3\) & 
        \(\mathrm{SU}(3)\times\mathrm{SU}(3)\times\mathrm{SU}(3)\) & \\ 
  \hline
  \(\mathrm{E}_7\) & \(\mathbbm{Z}_2\) & 
        \(\mathrm{SO}(12)\times\mathrm{SU}(2)\) & \\
  \(\mathrm{E}_7\) & \(\mathbbm{Z}_3\) & 
        \(\mathrm{SU}(6)\times\mathrm{SU}(3)\) & \\
  \(\mathrm{E}_7\) & \(\mathbbm{Z}_4\) & 
        \(\mathrm{SU}(4)\times\mathrm{SU}(4)\times\mathrm{SU}(2)\)
        & not maximal\\
  \(\mathrm{E}_7\) & \(\mathbbm{Z}_2\) & 
        \(\mathrm{E}_6\times\mathrm{U}(1)\) & \\        
  \(\mathrm{E}_7\) & \(\mathbbm{Z}_2\) & 
        \(\mathrm{SU}(8)\) & \\
  \hline
  \(\mathrm{E}_8\) & \(\mathbbm{Z}_2\) & 
        \(\mathrm{SO}(16)\) & \\
  \(\mathrm{E}_8\) & \(\mathbbm{Z}_4\) & 
        \(\mathrm{SU}(8)\times\mathrm{SU}(2)\) & not maximal\\
  \(\mathrm{E}_8\) & \(\mathbbm{Z}_6\) & 
        \(\mathrm{SU}(6)\times\mathrm{SU}(3)\times\mathrm{SU}(2)\)
        & not maximal\\ 
  \(\mathrm{E}_8\) & \(\mathbbm{Z}_5\) & 
        \(\mathrm{SU}(5)\times\mathrm{SU}(5)\) &\\
  \(\mathrm{E}_8\) & \(\mathbbm{Z}_4\) & 
        \(\mathrm{SO}(10)\times\mathrm{SU}(4)\) & not maximal\\
  \(\mathrm{E}_8\) & \(\mathbbm{Z}_3\) & 
        \(\mathrm{E}_6\times\mathrm{SU}(3)\) & \\
  \(\mathrm{E}_8\) & \(\mathbbm{Z}_2\) & 
        \(\mathrm{E}_7\times\mathrm{SU}(2)\) &\\
  \(\mathrm{E}_8\) & \(\mathbbm{Z}_3\) & 
        \(\mathrm{SU}(9)\) & \\ 
  \hline
 \end{tabular}
\end{center}
 \caption{Maximal subgroups of the simple groups and the corresponding
  \(\mathbbm{Z}_n\) orbifold twists. The five non-maximal subgroups which
  can be obtained by removing one node of the extended Dynkin diagram
  are listed for the sake of completeness.}
 \label{tab:OrbifoldTwists}
\end{table}


\begin{thebibliography}{99}

\bibitem{gg}
H. Georgi and S. Glashow, Phys. Rev. Lett. {\bf 32} (1974) 438.

\bibitem{gfm}
H.~Georgi, AIP Conf.\ Proc.\  {\bf 23} (1975) 575;\\
H.~Fritzsch and P.~Minkowski, Annals Phys.\  {\bf 93} (1975) 193.

\bibitem{ps}
J.~Pati and A.~Salam, Phys. Rev. {\bf D8} (1973) 1240;\\
J.~Pati and A.~Salam, Phys. Rev. {\bf D10} (1974) 275.

\bibitem{wy}
T.~Watari and T.~Yanagida, Phys.\ Lett.\ B {\bf 532} (2002) 252
[arXiv:hep-ph/0201086].

\bibitem{Babu:2002ti}
K.~S.~Babu, S.~M.~Barr and B.~s.~Kyae,
Phys.\ Rev.\ D {\bf 65} (2002) 115008\\{}
[arXiv:hep-ph/0202178].

\bibitem{nb}
G.~Burdman and Y.~Nomura,
Nucl.\ Phys.\ B {\bf 656} (2003) 3
[arXiv:hep-ph/0210257].

\bibitem{gmn}
I.~Gogoladze, Y.~Mimura and S.~Nandi,
arXiv:hep-ph/0304118.

\bibitem{wit}
E.~Witten, Nucl.\ Phys.\ B {\bf 258} (1985) 75.

\bibitem{dhvw}
L.~J.~Dixon, J.~A.~Harvey, C.~Vafa and E.~Witten, Nucl.\ Phys.\ B {\bf 261} 
(1985) 678 and {\bf 274} (1986) 285.

\bibitem{kaw}
Y.~Kawamura,
Prog.\ Theor.\ Phys.\  {\bf 105} (2001) 999
[arXiv:hep-ph/0012125].

\bibitem{af}
G.~Altarelli and F.~Feruglio, Phys.\ Lett.\ B {\bf 511} (2001) 257
[arXiv:hep-ph/0102301].

\bibitem{hn}
L.~J.~Hall and Y.~Nomura, Phys.\ Rev.\ D {\bf 64} (2001) 055003
[arXiv:hep-ph/0103125].

\bibitem{hm}
A.~Hebecker and J.~March-Russell, Nucl.\ Phys.\ B {\bf 613} (2001) 3\\{}
[arXiv:hep-ph/0106166].

\bibitem{abc} 
T.~Asaka, W.~Buchm\"uller and L.~Covi, Phys.\ Lett.\ B {\bf 523} (2001) 199
\\{} [arXiv:hep-ph/0108021]. 

\bibitem{hnos}
L.~J.~Hall, Y.~Nomura, T.~Okui and D.~R.~Smith, Phys.\ Rev.\ D {\bf 65} 
(2002) 035008 [arXiv:hep-ph/0108071].

\bibitem{dyn1}
E. B. Dynkin, \emph{The structure of semi-simple algebras}, 
Amer. Math. Soc. Transl. No. 1 (1950) pp. 1-143,\\ also
Amer. Math. Soc. Transl. (1) {\bf 9} (1962) pp. 328-469.

\bibitem{dyn2}
E. B. Dynkin, \emph{Semi-simple subalgebras of semi-simple Lie algebras},
Amer. Math. Soc. Transl., Ser. 2, {\bf 6} (1957) pp. 111-244.

\bibitem{dyn3}
E. B. Dynkin, \emph{Maximal subgroups of the classical groups},
Amer. Math. Soc. Transl., Ser. 2, {\bf 6} (1957) pp. 245-378.

\bibitem{dynb}
E. B. Dynkin, \emph{Selected papers}, Amer. Math. Soc., 1999. 

\bibitem{gil}
R.~Gilmore, \emph{Lie groups, Lie algebras, and some of their applications}, 
\\{} Malabar Krieger, 1994. 

\bibitem{cahn}
R.~N.~Cahn, \emph{Semisimple Lie algebras and their representations},\\{}
Benjamin/Cummings, 1984. 

\bibitem{Georgi:1999jb}
H.~Georgi, \emph{Lie algebras in particle physics. {F}rom isospin to unified
  theories, 2nd edition}, vol.~54, Perseus Books, 1999.

\bibitem{sla}
R.~Slansky, Phys.\ Rept.\  {\bf 79} (1981) 1.

\bibitem{hm1}
A.~Hebecker and J.~March-Russell, Nucl. Phys. \textbf{B625} (2002) 128
\\{}[arXiv:hep-ph/0107039].

\bibitem{Haba:2002py}
N.~Haba, M.~Harada, Y.~Hosotani and Y.~Kawamura,
Nucl.\ Phys.\ B {\bf 657} (2003) 169
[arXiv:hep-ph/0212035].

\bibitem{orb}
L.~J.~Hall and Y.~Nomura, arXiv:hep-ph/0212134;\\
M.~Quiros, arXiv:hep-ph/0302189.

\bibitem{Fegan:1991jb}
H.~D.~Fegan, \emph{An introduction to compact {L}ie groups}, World Scientific
  Publishing, 1991, 131 P.

\bibitem{dmr}
K.~R.~Dienes and J.~March-Russell,
Nucl.\ Phys.\ B {\bf 479} (1996) 113
\\{}[arXiv:hep-th/9604112].

\bibitem{kkkot}
Y.~Katsuki, Y.~Kawamura, T.~Kobayashi, N.~Ohtsubo and K.~Tanioka, Prog.\ 
Theor.\ Phys.\  {\bf 82} (1989) 171.\\
%
Y.~Katsuki, Y.~Kawamura, T.~Kobayashi, N.~Ohtsubo, Y.~Ono and K.~Tanioka,
Nucl.\ Phys.\ B {\bf 341} (1990) 611.

\bibitem{Golubitsky:1971ex}
M.~Golubitsky and B.~Rothschild, \emph{Primitive subalgebras of exceptional
  {L}ie algebras}, Pac. J. Math. \textbf{39 No. 2} (1971), 371--393.

\bibitem{abc1}
T.~Asaka, W.~Buchm\"uller and L.~Covi, Phys.\ Lett.\ B {\bf 540} (2002) 295
\\{}[arXiv:hep-ph/0204358].

\bibitem{Haba:2002vc}
N.~Haba and Y.~Shimizu,
arXiv:hep-ph/0212166.

\bibitem{Ibanez:1987xa}
L.~E.~Ibanez, H.~P.~Nilles and F.~Quevedo,
Phys.\ Lett.\ B {\bf 192} (1987) 332.

\bibitem{Ibanez:1986tp}
L.~E.~Ibanez, H.~P.~Nilles and F.~Quevedo,
Phys.\ Lett.\ B {\bf 187} (1987) 25.

\bibitem{Wigner}
E.~P.~Wigner, \emph{Group theory}, Academic press, 1949, 372 P.

\bibitem{GrootNibbelink:2003gd}
S.~Groot Nibbelink,
arXiv:hep-th/0305139.

\bibitem{nsw}
Y.~Nomura, D.~R.~Smith and N.~Weiner, Nucl.\ Phys.\ B {\bf 613} (2001) 147
\\{}[arXiv:hep-ph/0104041].

\bibitem{heb}
A.~Hebecker, Nucl.\ Phys.\ B {\bf 632} (2002) 101 [arXiv:hep-ph/0112230].

\bibitem{li}
T.~j.~Li, Nucl.\ Phys.\ B {\bf 633} (2002) 83 [arXiv:hep-th/0112255].

\bibitem{hos}
Y.~Hosotani, Phys.\ Lett.\ B {\bf 126} (1983) 309, and Annals Phys.\ 
{\bf 190} (1989) 233.

\bibitem{Candelas:en}
P.~Candelas, G.~T.~Horowitz, A.~Strominger and E.~Witten,
Nucl.\ Phys.\ B {\bf 258} (1985) 46.

\bibitem{mss}
N.~Marcus, A.~Sagnotti and W.~Siegel, Nucl.\ Phys.\ B {\bf 224} (1983) 159.

\bibitem{Imamura:2001es}
Y.~Imamura, T.~Watari and T.~Yanagida,
Phys.\ Rev.\ D {\bf 64} (2001) 065023
\\{}[arXiv:hep-ph/0103251].

\bibitem{abca}
T.~Asaka, W.~Buchm\"uller and L.~Covi, Nucl.\ Phys.\ B {\bf 648} (2003) 231
\\{}[arXiv:hep-ph/0209144].

\bibitem{ggnow}
F.~Gmeiner, S.~Groot Nibbelink, H.~P.~Nilles, M.~Olechowski and 
M.~G.~Walter, Nucl.\ Phys.\ B {\bf 648} (2003) 35 [arXiv:hep-th/0208146].

\bibitem{gq}
G.~von Gersdorff and M.~Quiros, arXiv:hep-th/0305024.

\bibitem{far}
A.~E.~Faraggi, talk at {\it 4th Int. Conf. on Phys. Beyond the Standard 
Model}, Lake Tahoe, 1994 [arXiv:hep-ph/9501288].

\bibitem{dm}
R.~Dermisek and A.~Mafi, Phys.\ Rev.\ D {\bf 65} (2002) 055002
[arXiv:hep-ph/0108139];\\
H.~D.~Kim and S.~Raby, JHEP {\bf 0301} (2003) 056 [arXiv:hep-ph/0212348].

\bibitem{kim1}
J.~E.~Kim, arXiv:hep-th/0301177;\\
K.~S.~Choi and J.~E.~Kim, arXiv:hep-ph/0305002.

\bibitem{kim}
K.~S.~Choi, K.~Hwang and J.~E.~Kim,
arXiv:hep-th/0304243.

\end{thebibliography}
\end{document}